\documentclass[a4paper,journall]{new-aiaa}

\usepackage[utf8]{inputenc}
\usepackage{textcomp}

\usepackage{graphicx}
\usepackage[labelformat=simple]{caption,subcaption}
\usepackage[english]{babel}
\usepackage{amsmath}
\usepackage{amssymb}
\usepackage{csvsimple}
\usepackage{array}
\usepackage[version=4]{mhchem}
\usepackage{multirow}

\newtheorem{myhyp}{Assumption}
\newtheorem{mydef}{Definition}
\newtheorem{paxmetric}{Proposed passenger-centric metric}
\newtheorem{paxmetric2}{Proposed passenger-centric metric}

\newcommand{\st}{$^\text{st}$}
\newcommand{\nd}{$^\text{nd}$}
\newcommand{\rd}{$^\text{rd}$}
\newcommand{\nth}{$^\text{th}$}

\graphicspath{{figures/}{figures/cbp/}{figures/features/}{figures/topics/}{figures/safegraph/}}

\title{Putting the Air Transportation System to sleep: a passenger perspective measured by passenger-generated data}
\author{Philippe Monmousseau \footnote{PhD student, Optimization and Machine Learning Team, philippe.monmousseau@enac.fr} and Daniel Delahaye\footnote{Professor, Head of Optimization and Machine Learning Team, delahaye@recherche.enac.fr}}
\affil{ENAC, Université de Toulouse, Toulouse, France}
\author{Aude Marzuoli\footnote{Principal Scientist, Replica, amarzuoli3@gatech.edu}}
\affil{Georgia Institute of Technology, Atlanta GA, USA}
\author{Eric Feron\footnote{Professor, Division of Electrical, Computer and Mathematical Science and Engineering, eric.feron@kaust.edu.sa}}
\affil{King Abdullah University of Science and Technology}
\begin{document}
\maketitle
\begin{abstract}
%%%%%
% Abstract ici
%
This paper aims at analyzing the effect on the US air transportation system of the travel restriction measures implemented during the COVID-19 pandemic from a passenger perspective. Flight centric data are not already publicly and widely available therefore the traditional metrics used to measure the state of this system are not yet available. Seven metrics based on three different passenger-generated datasets are proposed here. They aim to measure in close to real-time how the travel restriction measures impacted the relation between major stakeholders of the US air transportation system, namely passengers, airports and airlines. \\
\emph{Keywords - Air transportation system, passenger-generated data, passenger-centric metrics, COVID-19}
%
%%%%
\end{abstract}

%%%%%%
% Write content here
%

\section{Motivation}
\label{sec:intro}
\subsection{The COVID-19 pandemic and international travel from a US perspective}
\label{sec:intro_lock}
Following the outbreak of corona disease 2019 (COVID-19) caused by the sever acute respiratory syndrome coronavirus 2 (SARS-CoV-2), and to curb the spread of the resulting pandemic, many countries around the world have imposed travel restrictions, both domestically and internationally  \cite{NYT2020travelrestrictions}. 

Italy was the first country to implement a national lockdown situation \cite{WA2020lockdowns} on March 9\nth\,2020, previously only confining the northern region of Lodi on February 21\st\,2020, which was the center of its national COVID-19 outbreak. Two days later, on March 11\nth\,2020, the United States barred travellers who had visited China, Iran and 26 member states of the European Union (EU), and extended it to the other 2 members of the EU on March 16\nth\,2020 \cite{NYT2020travelrestrictions}. The EU then closed the boarders of 26 of its member states to nearly all visitors from non-EU countries on March 17\nth\,2020 \cite{NYT2020travelrestrictions}. And on March 19\nth\,2020, the US State Department issued a Level 4 Travel Advisory, a recommendation for all US citizens to avoid all international travel, still in place as of April 27\nth\,2020 \cite{TSG2020travelAdvisory}.

The following dates are represented in every graph presented in this paper using dotted lines in order to help better understand the timeline represented within each figure.
\begin{enumerate}
\item \textbf{Lodi} region lockdown in Italy: February 21\st, 2020
\item \textbf{Italy} lockdown: March 9\nth, 2020
\item \textbf{US ban} of EU travelers: March 11\nth, 2020
\item \textbf{EU} border \textbf{closure}: March 17\nth, 2020
\item \textbf{US Level 4} travel advisory: March 19\nth, 2020
\end{enumerate}

\subsection{What data can we use to assess the impact in real-time?}
The effect of these travel restrictions measures, and the other measures taken by a majority of countries worldwide, on the air transportation system has to be unprecedented. Official flight data are still to be released in the United States regarding international and domestic air transportation (as of April 27\nth\,2020), so there are no unified means of measuring this impact. 

Most traditional metrics to measure the state of the air transportation system are centered on the performance of flights in terms of delay, cancellation and number of passengers transported using data gathered by the Bureau of Transportation Statistics (BTS) \cite{BTSwebsite}. This data has first to be provided by airlines and airports to the BTS before being published as a monthly report. The usual latency is of two months for on-time flight data to be published by BTS. This frequency is not adapted to the monitoring of situations such as the COVID-19 pandemic.

This paper proposes to take an alternative approach, and to consider data generated by the core of the business of the air transportation system: passengers. Passengers generate various sort of data throughout their journey, as well as before and after their flight should have taken place. Some of these passenger-generated data are publicly available in close-to real-time and could be used in an aggregated and anonymized fashion to assess the state of the air transportation system. Three such data sources are considered in this paper in order to build passenger-centric metrics: data passively emitted by passenger passports at immigration, data passively emitted by their phones and data actively emitted within social media.

The rest of this paper is structured as follows: Section\,\ref{sec:background} first presents an overview of what has previously been done from a passenger perspective to monitor the state of the air transportation system. Section\,\ref{sec:airports} analyses the impact of the COVID-19 pandemic on airports from a passenger perspective. Section\,\ref{sec:airlines} then focuses on the impact on airlines and Section\,\ref{sec:concl} summarizes the passenger-centric metrics proposed and discusses future research direction.

\section{Background}
\label{sec:background}
A passenger approach to analyzing flight delays was first introduced by Bratu and Barnhart \cite{bratu2005AnalysisPassengerDelays} who developed a Passenger Delay Calculator to show that flight-centric metrics do not accurately reflect passenger delays, especially due to flight cancellations. Later in \cite{bratu2006FlightOperationsRecovery} they calculated passenger delay using monthly data from a major airline operating a hub-and-spoke network. They show that disrupted passengers, whose journey was interrupted by a capacity reduction, are only 3\% of the total passengers, but suffer 39\% of the total passenger delay.
Wang et al. in \cite{wang2006PassengerTripTime,wang2007MethodsAnalysisPassenger} showed that high passenger trip delays are disproportionately generated by canceled flights and missed connections. 9 of the busiest 35 airports cause 50\% of total passenger trip delays. Congestion, flight delay, load factor, flight cancellation time and airline cooperation policy are the most significant factors affecting total passenger trip delay.
These studies, based on BTS or airline data, have highlighted the disproportionate impact of airside disruptions on passenger door-to-door journeys, already showing that traditional flight-centric metrics do not capture the full picture.

This led NextGen \cite{nextgen} in the United States and ACARE Flightpath 2050 \cite{darecki2011Flightpath2050Europe} to advocate a shift from flight-centric metrics to passenger-centric metrics to evaluate the performance of the Air Transportation System.
Both the USA and Europe aim to take a more passenger-centric approach, with ACARE Flightpath 2050 setting some ambitious goals, including some that are not measurable yet due to lack of available data. In the US, the Joint Planning and Development Office has proposed and tested metrics regarding NextGen's goals, but there are still metrics missing from the passenger's viewpoint, especially regarding door-to-door travel times and passenger handling \cite{gawdiak2011NextGenMetricsJoint}.

The shift from flight-centric information to passenger-centric metrics was then explored by Cook et al. \cite{cook2012PassengerOrientedEnhancedMetrics} within the project POEM - Passenger Oriented Enhanced Metrics, where they designed propagation-centric and passenger-oriented performance metrics using both complexity and data science approaches. They simulated air transportation networks and analyzed their resilience from a flight-centric perspective and from a passenger-oriented perspective, highlighting the need for the implementation of passenger centric metrics.

Taking the passenger objectives during decision making was then proposed within the concept of Multimodal, Efficient Transportation in Airports and Collaborative Decision Making (META-CDM) by Laplace et al. \cite{laplace2014METACDMMultimodalEfficient}. This concept proposes to link both airside CDM and landside CDM, taking into account the passenger perspective. Within this framework, Kim et al. \cite{kim2013AirportGateScheduling} proposed to improve airport gate scheduling by implementing a decision model that balances aircraft, operator and passenger objectives. Dray et al. \cite{dray2015AirTransportationMultimodal} highlighted the need of taking a multi-modal approach, which is passenger centered, when handling major disturbances of the air transportation system in order to offer better solutions to passengers.

Taking a multi-modal approach implies having access to different source of data and being able to link them together. Data generated by passengers throughout there trip are diverse and scattered across different sensors. Airports gather customs or security records, shuttle traffic, parking occupancy, sometimes measure queue lengths, while third-parties collect online traces through WiFi hotspots and Bluetooth beacons \cite{sita}. These real-time information, combined with historical data, were used to analyze and predict passenger flow to an Australian immigration booth \cite{nikoue2015PassengerFlowPredictions} or within several Dutch train stations \cite{vandenheuvel2016AdvancesMeasuringPedestrians} as well as for the analysis and prediction of passenger occupancy in a Chinese airport \cite{huang2019ModelingPredictingOccupancy}. 
These studies are limited to a limited part of the full system (one or two airport terminals) indicating the difficulty of gathering a system-wide data-driven picture of passenger behavior. 

Considering passengers as sensors was made easier with the increase in the use of smartphones. Marzuoli et al. was able in \cite{marzuoli2019ImplementingValidatingAir} and \cite{marzuoli2018PassengercentricMetricsAir} to use mobile phone data in order to analyze the performances of US airports from a passenger perspective. These studies were a first validation that passenger-centric data can be used to have a view of the overall health of the Air Transportation System that is complementary to the traditional flight-centric approach. A major weather perturbation impact on passenger experience in airports was studied using this same approach and the study was complemented by an additional passenger generated data source, i.e. social media, in \cite{marzuoli2018PassengercentricMetricsAir}. In Europe, a similar approach was conducted within the BigData4ATM project\footnote{\url{www.bigdata4atm.eu}} by Garcia-Albertos et al. \cite{garcia-albertos2017UnderstandingDoortoDoorTravel} who were able to measure door-to-door travel times of air passengers between two Spanish cities, Madrid and Barcelona, thanks to mobile phone data. However mobile phone data is proprietary data and is not often publicly available for research. In the special case of research about the COVID-19 pandemic, SafeGraph \cite{SFGwebsite} gave access to an aggregated version of their database, consisting of various sets of data generated by mobile phone users. 

One important source of user-generated data regularly used to study large-scale behaviors with the advantage of real time availability is social media, and Twitter\footnote{\url{www.twitter.com}} more specifically. With more than 68 millions active users in the United States \cite{statTwitterUsers}, Twitter is an important pool of user-created data. Its real-time availability already led Twitter to be the main focus of multiple studies of large scale events, with several works by Palen et al. on how to help emergency responders during US natural disasters \cite{palen2010TwitterbasedInformationDistribution,vieweg2010MicrobloggingTwoNatural,kireyev2009ApplicationsTopicsModels}. 
In Europe, Terpstra et al. also studied how a real time Twitter analysis could provide valuable information for the operational response of a natural disaster crisis management with the case of the storm hitting a festival in Belgium \cite{terpstra2012RealtimeTwitterAnalysis}. 
Regarding the air transportation field, most works mining Twitter data focus on airline sentiment analysis, with Breen \cite{breen2012mining} explaining how to mine Twitter textual data and create sentiment classifiers or Wang et al. \cite{wan2015EnsembleSentimentClassification} proposing an improved airline sentiment classifying method. These works focused essentially on improving the available methods for sentiment analysis without proposing any direct use of their results to improve airline service or passenger satisfaction.
Monmousseau et al. in \cite{monmousseau2019PredictingAnalyzingUS} used publicly available social media data created by passengers to accurately estimate and predict the hourly aggregated status of the US air transportation system. This method was further improved in \cite{monmousseau2019PassengersSocialMediaa} to reliably estimate the hourly delays at departure and at arrival per airport.

\section{Impact the COVID-19 travel restriction measures on airports}
\label{sec:airports}
This section leverages two different user-generated datasets to analyze what was the effect on US airports of the implemented travel restrictions presented in Section\,\ref{sec:intro_lock} from a passenger and visitor perspective. 

\subsection{Overall impact on the number of passengers/visitors at airports}
\subsubsection{International passengers}
Travel restrictions do not ban entirely international travel, and there are still passengers arriving at most US airports of entry after the implementation of these travel restrictions. However, starting March 13\nth\,2020, US citizens who have been in high risk areas and are returning to the United States have to arrive by one the thirteen following airports of entry: \cite{NYT2020travelrestrictions}
\begin{itemize}
\item ATL: Hartsfield-Jackson Atlanta International Airport
\item BOS: Boston-Logan International Airport
\item DFW: Dallas Fort Worth International Airport
\item DTW: Detroit Metropolitan Airport
\item EWR: Newark Liberty International Airport
\item HNL: Daniel K. Inouye International Airport 
\item IAD: Washington-Dulles International Airport
\item JFK: John F. Kennedy International Airport 
\item LAX: Los Angeles International Airport
\item MIA: Miami International Airport
\item ORD: Chicago O'Hare International Airport
\item SEA: Seattle-Tacoma International Airport
\item SFO: San Francisco International Airport
\end{itemize}

The effect of these travel restrictions on international travel coming to the US can be studied thanks to the "Airport Wait Times" data from the Customs and Border Protection (CBP) website \cite{CBPwebsite}. This data are aggregated at an hourly level and are usually available on the following day they are generated. The readiness of the data is due to the fact that CBP measures directly the signal emitted by passengers, the signal here being emitted through passports once passengers clear the immigration process, and does not have to wait for an airline or airport to process and provide the data. 

Among other information, the dataset contains the number of passengers arriving at immigration per hour, the average wait time at immigration per hour, and the number of open immigration booths per hour. For a more detailed presentation of the available dataset, the authors recommend the reading of \cite{monmousseau2019DoorwayUnitedStates}, which also proposes an analysis of these wait times from January 2013 to January 2019. The data considered here ranges from January 1\st\,2020 to April 22\nd\,2020.

\begin{figure}[ht]
\begin{center}
\includegraphics[width=\textwidth]{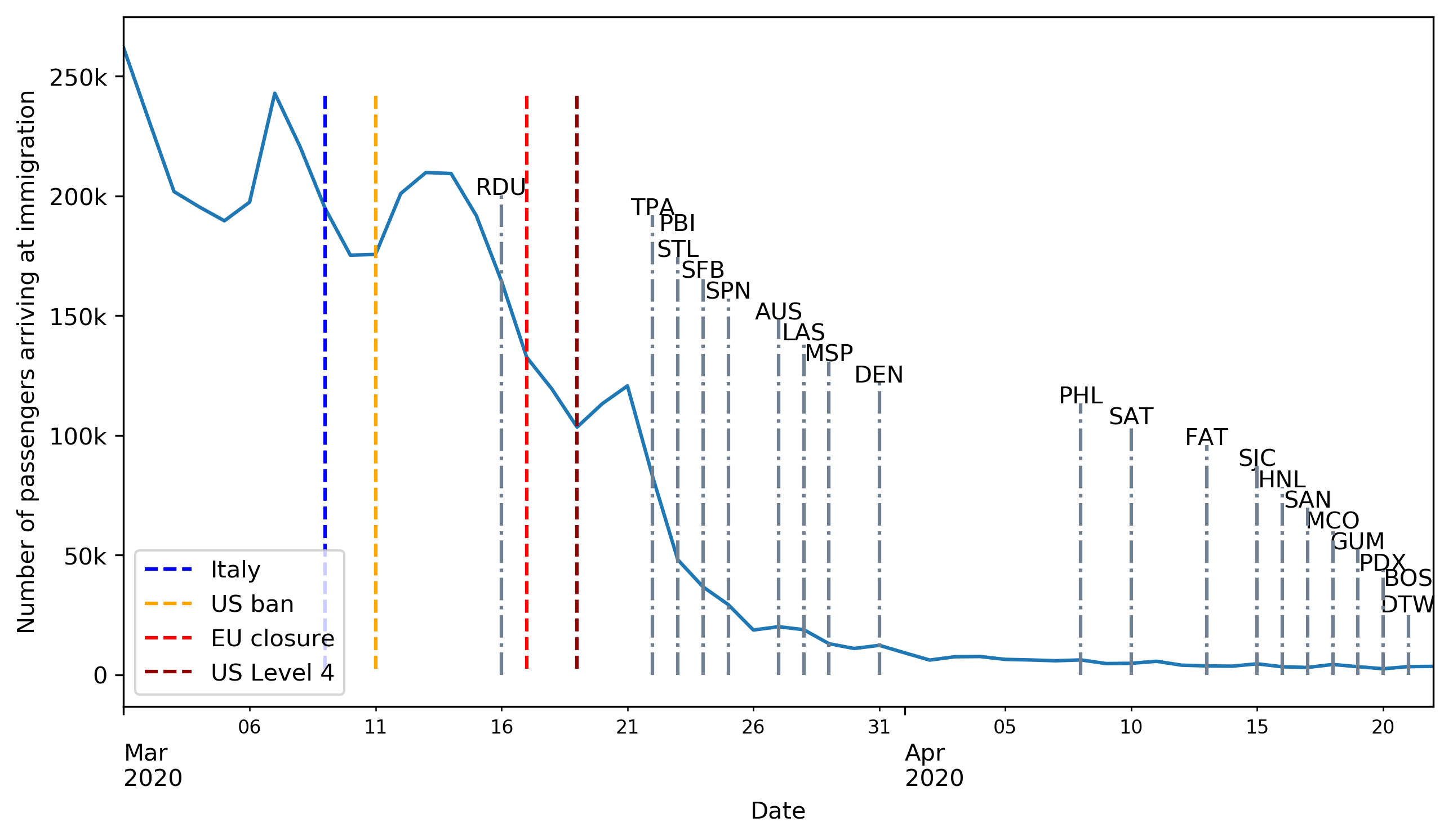}
\caption{Evolution of the total number of daily arriving passengers. The dates of last recorded CBP data for airports with no immigration data on the date of April 22\nd\,2020 are indicated as dotted lines.}
\label{fig:apt_closures}
\end{center}
\end{figure}

Looking first at the evolution of the total number of passengers arriving at US immigration booths per day across all airports, this number of passengers drops from an average of 218.7 thousands passengers per day between February 23\rd\,2020 and March 15\nth\,2020 to an average of only 5.0 thousands passengers per day between April 1\st\,2020 and April 22\nd\,2020. This represents a drop of 97.7\% in two weeks. The day by day evolution of the total number of passengers arriving at US immigration from March 1\st\,to April 22\nd\, is shown in Figure\,\ref{fig:apt_closures}. This figure also indicates for each airport with no CBP immigration data available on April 22\nd\,the last date where immigration data are available. This corresponds to 22 airports. Only Raleigh–Durham International Airport (RDU) closed its immigration service between the US ban of EU travellers and before the US entered a Level 4 travel advisory. It is to be noted that John Wayne Airport (SNA) has no immigration data since January 5\nth\,2020. Starting March 22\nd\,2020, the number of airports not generating any immigration data steadily increases with nine airports shutting down their immigration services in ten days. Another nine airports then stop generating immigration data in ten days starting April 12\nth\,2020.

Figure\,\ref{fig:apt_closures} indicates that BOS does not have any immigration data on April 22\nd\,2020 even though it is one of the selected airport of entry for US citizens coming from high-risk areas. This illustrates the fact that the influx of international passengers is so low that they had no arriving international passenger to their immigration service for at least one day.

\subsubsection{Impact on the number of airport visitors}
From a domestic perspective, and thanks to SafeGraph's willingness to provide aggregated data for research on how to better understand and better handle the COVID-19 pandemic, weekly patterns at specific points of interest (POI) are available\footnote{\url{https://docs.safegraph.com/docs/weekly-patterns}}. From these patterns, it is possible to have an estimate of the number of airport visitors per hour by considering all available POI associated with an airport. Airport visitors are a broader category than air passengers, since this category also encompasses airport staff and people dropping off or picking up passengers. The data available for this study ranges from February 27\nth\,2020 to April 18\nth\,2020.

\begin{figure}[ht]
\begin{center}
\includegraphics[width=\textwidth]{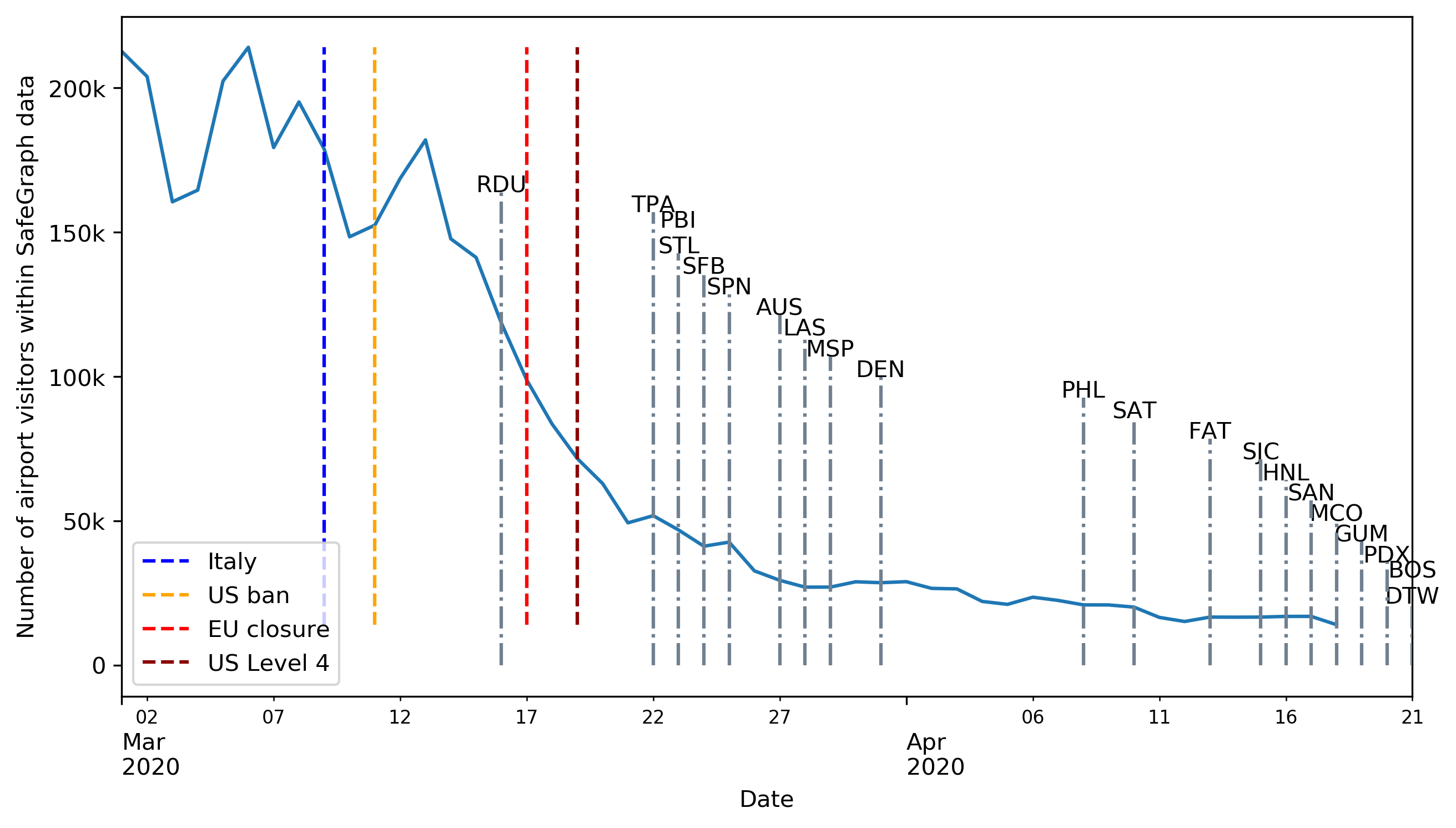}
\caption{Evolution of the total number of daily airport visitors using SafeGraph data. The dates of last recorded CBP data for airports with no immigration data on the date of April 22\nd\,2020 are indicated as dotted lines.}
\label{fig:apt_visitors}
\end{center}
\end{figure}

Looking first at the evolution of the total number of airport visitors per day across all airports, this number of passengers drops from an average of 176.8 thousands passengers per day between February 27\nth\,2020 and March 15\nth\,2020 to an average of only 20.2 thousands passengers per day between April 1\st\,2020 and April 18\nth\,2020. This represents a drop of 88.6\% in two weeks. The day by day evolution of the total number of airport visitors from March 1\st\,to April 18\nth\, is shown in Figure\,\ref{fig:apt_visitors}. Similarly to Figure\,\ref{fig:apt_closures}, this figure also indicates for each airport with no CBP immigration data available on April 22\nd\,the last date where immigration data are available.

From Figure\,\ref{fig:apt_visitors}, it is clear that US domestic travel was already impacted before the rise to a Level 4 travel advisory: The number of airport visitors contained within the SafeGraph data drops from 152.4 thousands on March 11\nth\,2020 down to 71.6 thousands on March 19\nth\,2020, which represents a 53\% drop.

\subsection{Distribution of the impact across airports}
\subsubsection{At immigration}
\begin{figure}[h!t]
\begin{center}
\begin{subfigure}[b]{.49\textwidth}
\includegraphics[width=\textwidth]{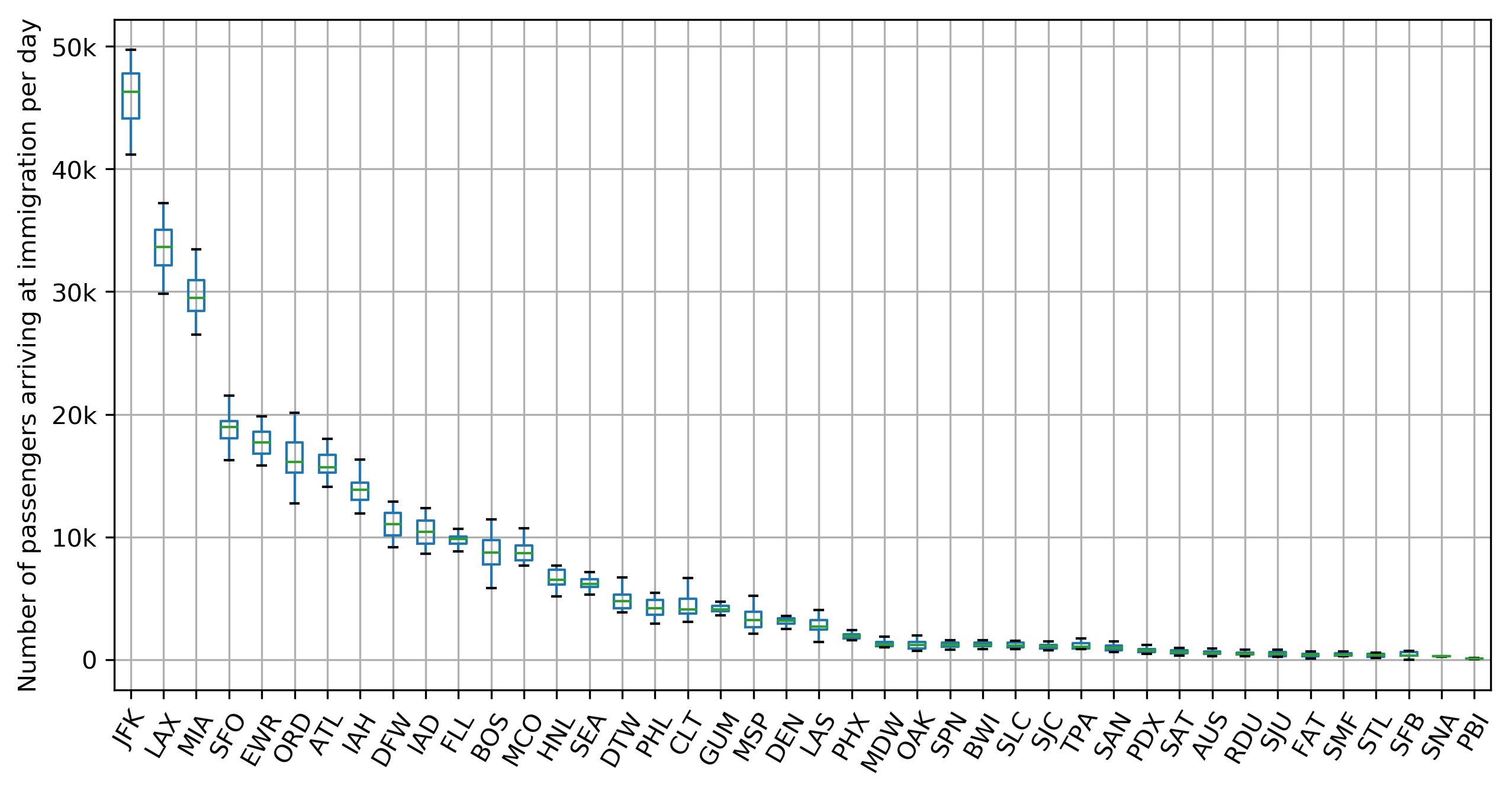}
\caption{April 1\st-22\nd\, 2019}
\label{fig:cbp_box_num19}
\end{subfigure}\hfill
\begin{subfigure}[b]{.49\textwidth}
\includegraphics[width=\textwidth]{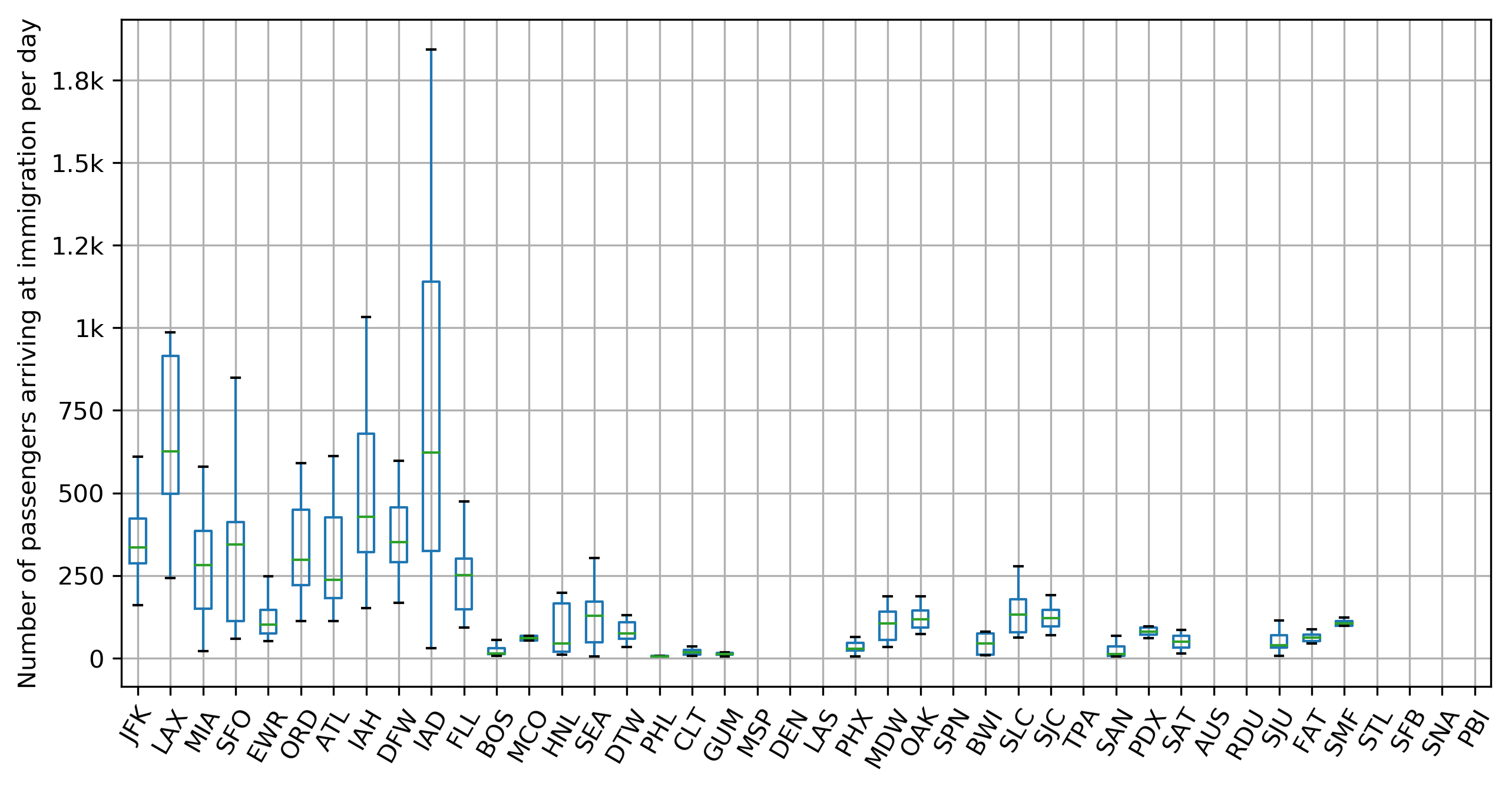}
\caption{April 1\st-22\nd\, 2020}
\label{fig:cbp_box_num20}
\end{subfigure}
\caption{Boxplots of the number of arriving passengers per day for each airport of entry to the US over the first three weeks of April for the years 2019 and 2020.}
\label{fig:cbp_box_num}
\end{center}
\end{figure}

Figure\,\ref{fig:apt_closures} showed that all US airports of entry are impacted by these lockdown and travel ban measures, and the next step is now to look into this impact at an airport level. Figure\,\ref{fig:cbp_box_num} compares the individual airport situation of the first two weeks of April 2020 with the first two weeks of April 2019. Figure\,\ref{fig:cbp_box_num19} shows the boxplots of the number of passengers arriving at immigration per day for each airport over the period of April 1\st-22\nd\,2019. The median number of arriving passengers is indicated in green and each box lower and upper bounds represent respectively the 1\st\,and 3\rd\, quartile. The whiskers above and below each box give a visualization of the full range of the considered data even though extreme values are not drawn. The airports are ordered by their median daily number of passengers arriving at immigration over that period. 
Figure\,\ref{fig:cbp_box_num20} shows the boxplots of the number of passengers arriving at immigration per day for each airport over the period of April 1\st-22\nd\,2020. The airports in this figure are in the same order as for Figure\,\ref{fig:cbp_box_num19}. Please note that the y-axis are not the same between Figure\,\ref{fig:cbp_box_num19} and Figure\,\ref{fig:cbp_box_num20} due to the important drop in the number of passengers arriving at US airports of entries after the implementation of the travel restriction measures.

Figure\,\ref{fig:cbp_box_num19} is a snapshot of the "normal" situation regarding the number of passengers arriving at US immigration over the first three weeks of April, while Figure\,\ref{fig:cbp_box_num20} is a snapshot of a pandemic situation regarding the number of passengers arriving at US immigration. The thirteen airports chosen for handling the return of US citizens from high-risk areas are all in the top 16 airports with the highest median daily number of passengers arriving at immigration, along with George Bush Intercontinental Airport (IAH), Fort Lauderdale–Hollywood International Airport (FLL) and Orlando International Airport (MCO).

The drop in the number of passengers arriving at immigration per day is clearly visible between the years 2019 (Figure\,\ref{fig:cbp_box_num19}) and 2020 (Figure\,\ref{fig:cbp_box_num20}) for the airports with the most arriving passengers. The drop for JFK is the most important in volume going from a median number of passengers arriving at immigration of 45.9 thousands between April 1\st-22\nd\, 2019 down to a median of 360 April 1\st-22\nd\, 2020. For JFK, this drop represents a drop of 99.3\% between the median number of passengers arriving at immigration of these two periods. For all the considered airports, the corresponding drop is between 70.7\% for Sacramento International Airport (SMF) and 100\% for the eleven airports without no immigration data between April 1\st\,2020 and April 22\nd\,2020.

Looking at the airport ranking based on the median number of passengers arriving at immigration per day over the period of April 1\st-22\nd, Figure\,\ref{fig:cbp_box_num20} shows that it has been reshuffled from year 2019 to year 2020: JFK dropped to the sixth place and IAD climbed to the first place right behind LAX. IAD has however the highest average number of passengers arriving at immigration per day over the period of April 1\st-22\nd\,2020 with 726 passengers a day on average, LAX being second with 658 passengers a day on average. JFK was the airport with the highest number of passengers arriving at immigration per year since 2013 \cite{CBPwebsite}, this makes the drop from first place to sixth place is all the more impressive.

\subsubsection{At airports}
\begin{figure}[h!t]
\begin{center}
\begin{subfigure}[b]{.49\textwidth}
\includegraphics[width=\textwidth]{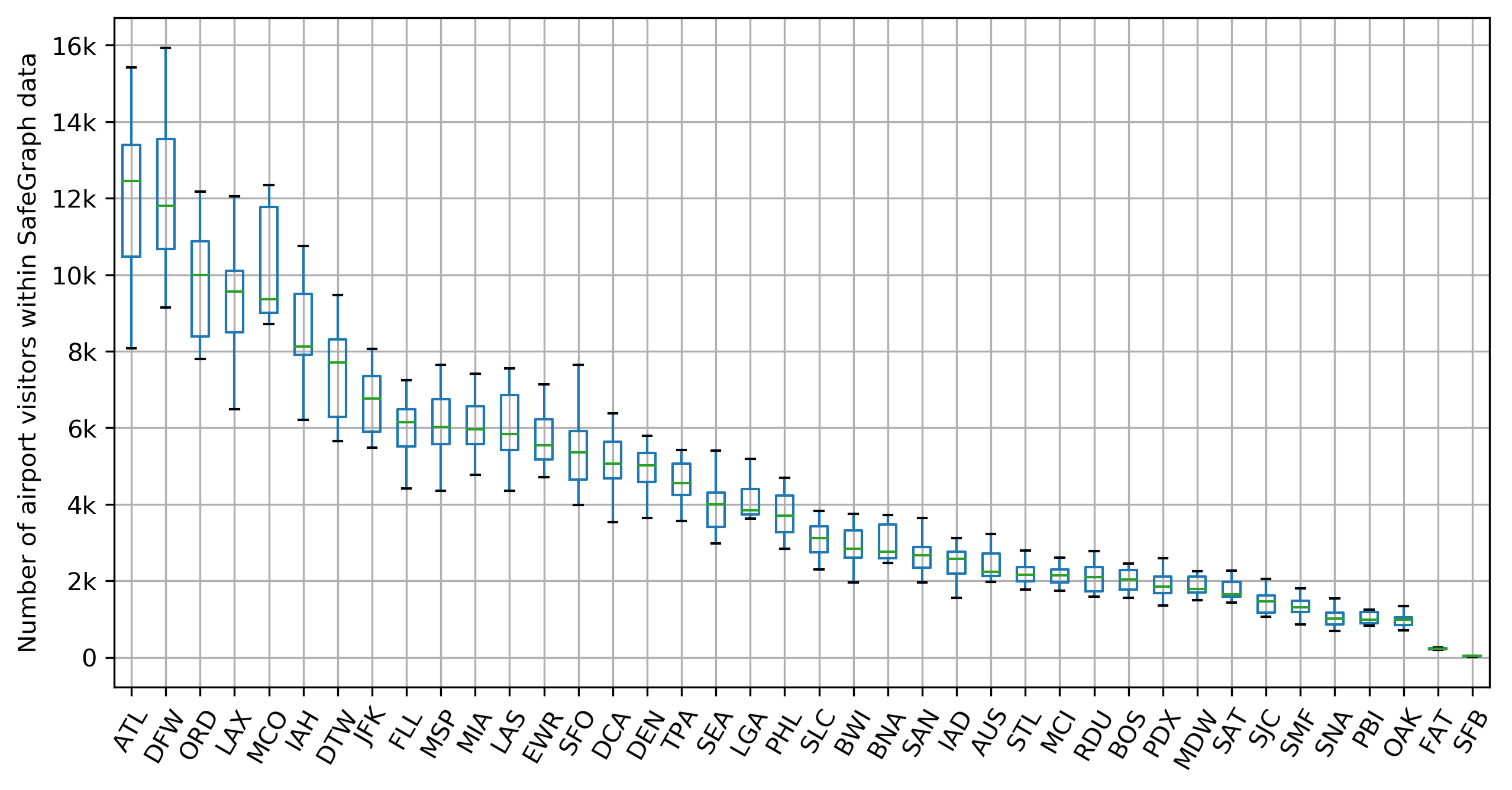}
\caption{March 1\st-15\nth 2020}
\label{fig:sfg_box_num19}
\end{subfigure}\hfill
\begin{subfigure}[b]{.49\textwidth}
\includegraphics[width=\textwidth]{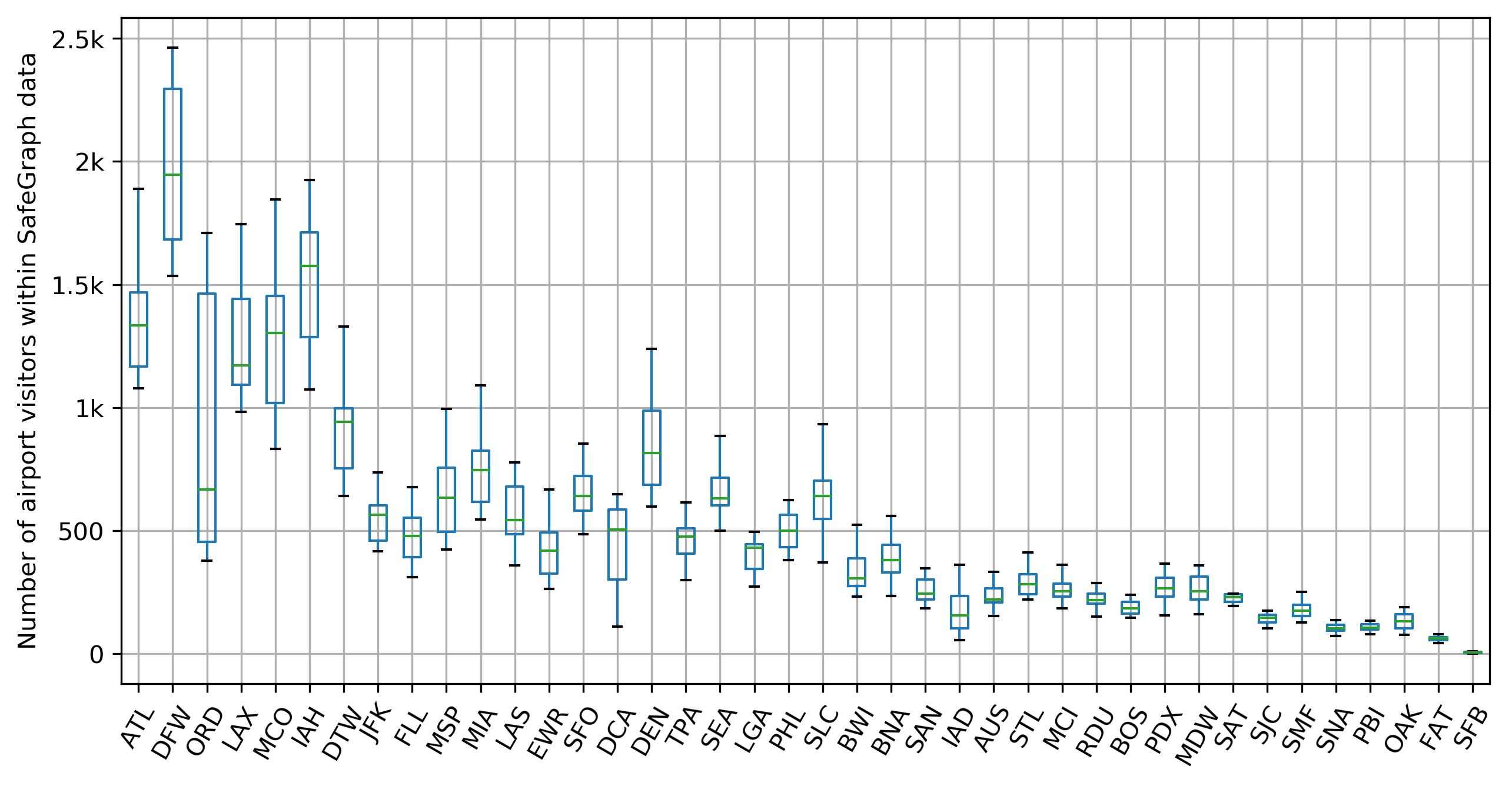}
\caption{April 1\st-15\nth 2020}
\label{fig:sfg_box_num20}
\end{subfigure}
\caption{Boxplots of the number of airport visitors per day for 44 US airport with available SafeGraph data over the first two weeks of March and April 2020.}
\label{fig:sfg_box_num}
\end{center}
\end{figure}

A similar comparison of the number of airport visitors before and after the travel restriction measures can be conducted based on the SafeGraph data. Due to data availability, this comparison has to take place between March 2020 and April 2020.
Figure\,\ref{fig:sfg_box_num} shows the boxplots of the number of airport visitors per day for 40 US airport with available SafeGraph data over the first two weeks of March 2020 (Figure\,\ref{fig:sfg_box_num19}) and April 2020 (Figure\,\ref{fig:sfg_box_num19}). The airports on these two plots are sorted by their median daily number of airport visitor over the period of March 1\st-15\nth\,2020. Please note that the y-axis are not the same between Figure\,\ref{fig:sfg_box_num19} and Figure\,\ref{fig:sfg_box_num20} due to the important drop in the number of passengers arriving at US airports of entries after the implementation of the travel restriction measures.
 
As for the number of passengers arriving at immigration, the airport with the highest median daily number of airport visitors is the airport with the most important drop in volume. Regarding the median daily number of airport visitors, ATL has a drop of 11.1 thousand airport visitors in the SafeGraph data between these two weeks, which represents a 89.3\% drop. Unlike for immigration, no airport stops completely of receiving visitors, though the drop is important for all 40 considered airports, ranging from 72.5\% for Fresno Yosemite International Airport (FAT) to 93.8\% for IAD.

\subsection{Proposed passenger-centric metrics}
\subsubsection{At immigration}
\label{sec:pax_immigration}
With 35 airports having a drop in the median number of passengers arriving at immigration greater than 90\%, all airports are severely impacted by the COVID-19 measures from a passenger-volume perspective. A question to be asked is: Since there are far fewer passengers arriving at immigration, does the immigration process go faster? The number of agents operating immigration booths has also decreased due to the corona virus, but it is possible to consider an immigration load factor. 
\begin{mydef}
\label{def:load_factor}
The \textbf{immigration load factor} is defined as the ratio of the number of passengers arriving at immigration per hour with the number of open immigration booths per hour. 
\end{mydef}
This load factor indicates the load in terms of passengers for each immigration booth per hour. A lower load indicates that each immigration booth has fewer passengers to process per hour. From a passenger perspective, a lower load for a given number of passengers, indicates that there are more immigration booths open, so the average processing time should be lower and thus a passenger at immigration would have to wait less to be processed.

\begin{myhyp}
\label{hyp:ratio_wait}
If the daily average immigration load factor decreases, then the daily average wait time for passengers at immigration should decrease as well.
\end{myhyp}

Based on this reasoning, an \textbf{immigration quality score} is proposed: It measures how well Assumption\,\ref{hyp:ratio_wait} is verified for an airport immigration service over a selected period of days. 

\begin{paxmetric}
\label{pax:immigration}
The \textbf{immigration quality score} for an airport of entry is defined as the correlation between the daily average wait time for passengers at its immigration service and the daily average immigration load factor of the airport over a given period.
\end{paxmetric}
This immigration quality score is equal to 1 if Assumption\,\ref{hyp:ratio_wait} is perfectly verified, to 0 if the daily average wait time for passengers at immigration is uncorrelated with the daily average immigration load factor and to -1 if the opposite of Assumption\,\ref{hyp:ratio_wait} occurs over the considered period, i.e. a decrease in the daily average immigration load factor implies an increase in the daily average wait time for passengers at immigration.

This proposed passenger-centric metric is applied to the period pre-COVID (January 1\st\,2020 to February 29\nth\,2020) and to the period post-COVID (March 1\st\,2020 to April 22\nd\,2020) for 40 US airports of entry. Table\,\ref{tab:cbp_ratio_ranking} shows the associated partial ranking (top ten best airports and top 10 worst airports) for these two periods.
\begin{table}[ht]
\begin{center}
\caption{Airport partial ranking based on the Proposed passenger-centric metric\,\ref{pax:immigration} (immigration quality score) applied to the period of pre-COVID of January 1\st\,2020 to February 29\nth\,2020 and to the period post-COVID of March 1\st\,2020 to April 22\nd\,2020 for the 40 considered US airports of entry.}
\label{tab:cbp_ratio_ranking}
\begin{tabular}{|c|c|c|c|c||c|c|c|c|c|}
\hline
\multicolumn{5}{|c||}{\bfseries Top ten best airports} & \multicolumn{5}{c|}{\bfseries Top ten worst airports} \\ 
\multirow[b]{2}{*}{\bfseries Rank} & \multicolumn{2}{c|}{\bfseries Pre-COVID} & \multicolumn{2}{c||}{\bfseries Post-COVID} & \multirow[b]{2}{*}{\bfseries Rank} & \multicolumn{2}{c|}{\bfseries Pre-COVID} & \multicolumn{2}{c|}{\bfseries Post-COVID} \\ 
\bfseries  & \bfseries Airport & \bfseries Score & \bfseries Airport & \bfseries Score & \bfseries  & \bfseries Airport & \bfseries Score & \bfseries Airport & \bfseries Score 
 \\ \hline
1&SFB&0.92&SFB&0.98&40&JFK&-0.32&SLC&-0.1\\
2&SPN&0.91&LAS&0.93&39&ATL&-0.17&AUS&0.17\\
3&PBI&0.72&SFO&0.89&38&MDW&-0.17&PDX&0.17\\
4&RDU&0.72&MSP&0.85&37&PDX&-0.09&HNL&0.17\\
5&GUM&0.68&PHX&0.84&36&OAK&-0.05&IAD&0.19\\
6&CLT&0.66&MIA&0.83&35&IAH&-0.01&BWI&0.21\\
7&SAT&0.63&DEN&0.82&34&LAX&0.02&MDW&0.25\\
8&MCO&0.62&FAT&0.82&33&MSP&0.02&SJU&0.29\\
9&TPA&0.61&CLT&0.81&32&SFO&0.12&OAK&0.34\\
10&PHL&0.6&JFK&0.81&31&FAT&0.14&DFW&0.4\\

\hline
\end{tabular}
\end{center}
\end{table}

The airport still generating immigration data on April 24\nth\, with the worst drop between the period pre-COVID and the period post-COVID is IAD, going from 17\nth\, down to 36\nth, and the airport still generating immigration data on April 24\nth\, with the best increase in rank is JFK, with 30 places gained and with an increase in score from the negative value of -0.32 to the positive value of +0.81.

\subsubsection{In terms of airport visitors}
With 38 airports having a drop in the median number of airport visitors greater than 80\%, all airports are also severely impacted by the COVID-19 travel restrictions measures from a visitor-volume perspective. Visitors in general avoid airports, but some are still going to the airports after the travel restriction measures. The same question as for the immigration process can be asked: Are these visitors processed faster since there are less visitors?

The data for visitors available for this is different than the data available for passengers arriving at immigration, therefore a different approach has to be considered here. The SafeGraph data contains weekly bucketed dwell times for each considered location. The dwell time is the time spent at that location, be it waiting, shopping, walking, etc. The buckets are: less than 5 minutes, between 5 and 20 minutes, between 21 and 60 minutes, between 61 minutes and 240 minutes and more than 240 minutes. From these weekly bucketed dwell times, two complementary passenger-metrics are proposed to measure an airport efficiency to process visitors.

\begin{paxmetric}
\label{pax:prop60}
The weekly \textbf{airport visitor efficiency score} for an airport is defined as the weekly proportion of airport visitors spending less than 60 minutes at an airport.
\end{paxmetric}

\begin{paxmetric}
\label{pax:prop240}
The weekly \textbf{airport visitor slugginess score} for an airport is defined as the weekly proportion of airport visitors spending more than 240 minutes at an airport.
\end{paxmetric}

The time limits within these two metrics are also chosen due to the format of the data, and could be adjusted to less aggregated data. The idea behind the airport visitor efficiency score is to incentivize airports to keep the flow of people coming in and out of their airports as fast as possible. The time limit of 60 minutes concerns essentially visitors dropping off or picking up a passenger, and hopefully some passengers on domestic flights, where the overall security screening process is faster than for international flights. Most airlines and airports recommend their passengers on international flights to arrive two to three hours ahead of their flight's scheduled departure time, therefore the idea behind the airport visitor slugginess score is to measure the validity of this recommendation.

Airport staff can be counted as airport visitors using this dataset and they are likely to stay more than 240 minutes at the airport, increasing the number of airport visitors staying longer than this threshold. Therefore, an airport with a high airport visitor slugginess score could either be an airport with many passengers taking more than four hours to clear their entire airport process, or an airport with a disproportionate number of airport staff compared to the number of airport visitors.

Since there are several locations per airport within the SafeGraph data, e.g. "LAX Terminal 4" and "LAX Terminal South" for LAX, an estimation of the proposed airport visitor efficiency score is calculated by taking the minimum weekly proportion of airport visitors spending less than 60 minutes at a location within an airport over all considered airport locations. Similarly, an estimation of the proposed airport visitor slugginess score is calculated by taking the maximum weekly proportion of airport visitors spending more than 240 minutes at a location within an airport over all considered airport locations.

These proposed passenger-centric metrics are applied to the period pre-COVID (March 1\st\,2020 to March 15\nth\,2020) and to the period post-COVID (April 5\nth\,2020 to April 19\nth\,2020) for 44 US airports. These periods contain 2 weeks each and therefore 2 points of data each. The scores are calculated for each week and then averaged over the period. Table\,\ref{tab:sfg_60_ranking} shows the partial ranking (top ten best airports and top 10 worst airports) associated to the proposed passenger-metric\,\ref{pax:prop60} for these two periods.
\begin{table}[h!t]
\begin{center}
\caption{Airport partial ranking using the proposed metric based on the proportion of airport visitors staying less than 60 minutes applied to the period of pre-COVID of March 1\st\,2020 to March 15\nth\,2020 and to the period post-COVID of April 5\nth\,2020 to April 19\nth\,2020 for the 44 considered US airports based on SafeGraph data.}
\label{tab:sfg_60_ranking}
\begin{tabular}{|c|c|c|c|c||c|c|c|c|c|}
\hline
\multicolumn{5}{|c||}{\bfseries Top ten best airports} & \multicolumn{5}{c|}{\bfseries Top ten worst airports} \\ 
\multirow[b]{2}{*}{\bfseries Rank} & \multicolumn{2}{c|}{\bfseries Pre-COVID} & \multicolumn{2}{c||}{\bfseries Post-COVID} & \multirow[b]{2}{*}{\bfseries Rank} & \multicolumn{2}{c|}{\bfseries Pre-COVID} & \multicolumn{2}{c|}{\bfseries Post-COVID} \\ 
\bfseries  & \bfseries Airport & \bfseries Score & \bfseries Airport & \bfseries Score & \bfseries  & \bfseries Airport & \bfseries Score & \bfseries Airport & \bfseries Score 
 \\ \hline
1&SJC&0.69&SJC&0.6&44&LGA&0.0&LAX&0.0\\
2&GUM&0.69&SMF&0.56&43&LAX&0.0&SLC&0.0\\
3&MCI&0.66&RDU&0.55&42&SFB&0.1&MIA&0.12\\
4&RDU&0.65&SAT&0.55&41&DEN&0.13&DEN&0.14\\
5&SMF&0.64&AUS&0.53&40&MIA&0.21&SFO&0.16\\
6&OAK&0.63&OAK&0.53&39&ATL&0.24&BNA&0.18\\
7&SAT&0.62&MCI&0.53&38&EWR&0.24&ATL&0.19\\
8&AUS&0.62&GUM&0.5&37&SLC&0.25&DTW&0.19\\
9&TPA&0.61&PHX&0.49&36&DCA&0.31&IAD&0.2\\
10&FAT&0.6&STL&0.48&35&PDX&0.32&EWR&0.2\\

\hline
\end{tabular}
\end{center}
\end{table}

In Table\,\ref{tab:sfg_60_ranking}, a score of 1 indicates that all airport visitors within the SafeGraph data spend less than one hour at the same location within the airport, while a score of 0 indicates that all airport visitors within the SafeGraph data spend more than one hour at the same location within the airport. Some airports have a score of 0 due to locations receiving very few visitors (less than 5) over the considered week that were captured within the SafeGraph data, and all those visitors stayed more than one hour at that same airport location.

Table\,\ref{tab:sfg_240_ranking} shows the partial ranking (top ten best airports and top 10 worst airports) associated to the proposed passenger-metric\,\ref{pax:prop240} for the same two considered periods.
\begin{table}[h!t]
\begin{center}
\caption{Airport partial ranking using the proposed metric based on the proportion of airport visitors staying more than 240 minutes applied to the period of pre-COVID of March 1\st\,2020 to March 15\nth\,2020 and to the period post-COVID of April 5\nth\,2020 to April 19\nth\,2020 for the 44 considered US airports based on SafeGraph data.}
\label{tab:sfg_240_ranking}
\begin{tabular}{|c|c|c|c|c||c|c|c|c|c|}
\hline
\multicolumn{5}{|c||}{\bfseries Top ten best airports} & \multicolumn{5}{c|}{\bfseries Top ten worst airports} \\ 
\multirow[b]{2}{*}{\bfseries Rank} & \multicolumn{2}{c|}{\bfseries Pre-COVID} & \multicolumn{2}{c||}{\bfseries Post-COVID} & \multirow[b]{2}{*}{\bfseries Rank} & \multicolumn{2}{c|}{\bfseries Pre-COVID} & \multicolumn{2}{c|}{\bfseries Post-COVID} \\ 
\bfseries  & \bfseries Airport & \bfseries Score & \bfseries Airport & \bfseries Score & \bfseries  & \bfseries Airport & \bfseries Score & \bfseries Airport & \bfseries Score 
 \\ \hline
1&GUM&0.04&GUM&0.0&44&LGA&1.0&LAX&1.0\\
2&MCI&0.05&SMF&0.16&43&SFB&0.73&SFB&0.71\\
3&SMF&0.05&SJC&0.17&42&DEN&0.62&DEN&0.67\\
4&SJC&0.05&HNL&0.19&41&SLC&0.52&SFO&0.55\\
5&AUS&0.05&RDU&0.2&40&EWR&0.49&LGA&0.55\\
6&OAK&0.05&PHX&0.2&39&MSP&0.48&ATL&0.53\\
7&STL&0.05&OAK&0.2&38&DTW&0.35&SLC&0.5\\
8&SAT&0.06&AUS&0.21&37&LAX&0.27&EWR&0.49\\
9&PHX&0.06&CLT&0.21&36&DFW&0.27&SNA&0.48\\
10&PBI&0.06&SAT&0.21&35&ATL&0.27&MDW&0.47\\

\hline
\end{tabular}
\end{center}
\end{table}

In Table\,\ref{tab:sfg_240_ranking}, a score of 0 indicates that all airport visitors within the SafeGraph data spend less than four hours at the same location within the airport, while a score of 1 indicates that all airport visitors within the SafeGraph data spend more than four hours at the same location within the airport. Similarly to the visitor airport efficiency score, some airports have a score of 1 due to locations receiving very few visitors (less than 5) over the considered week that were captured within the SafeGraph data, and all those visitors stayed more than four hours at that same airport location.

\subsection{Focus on JFK and IAD immigration process}
Following the results of the metric proposed in Section\,\ref{sec:pax_immigration}, this section focuses on the two airports with the most important change in behavior linked to the COVID-19 travel restriction measures, JFK and IAD.

\subsubsection{JFK}
\begin{figure}[h!t]
\begin{center}
\begin{subfigure}[t]{.49\textwidth}
\includegraphics[width=\textwidth]{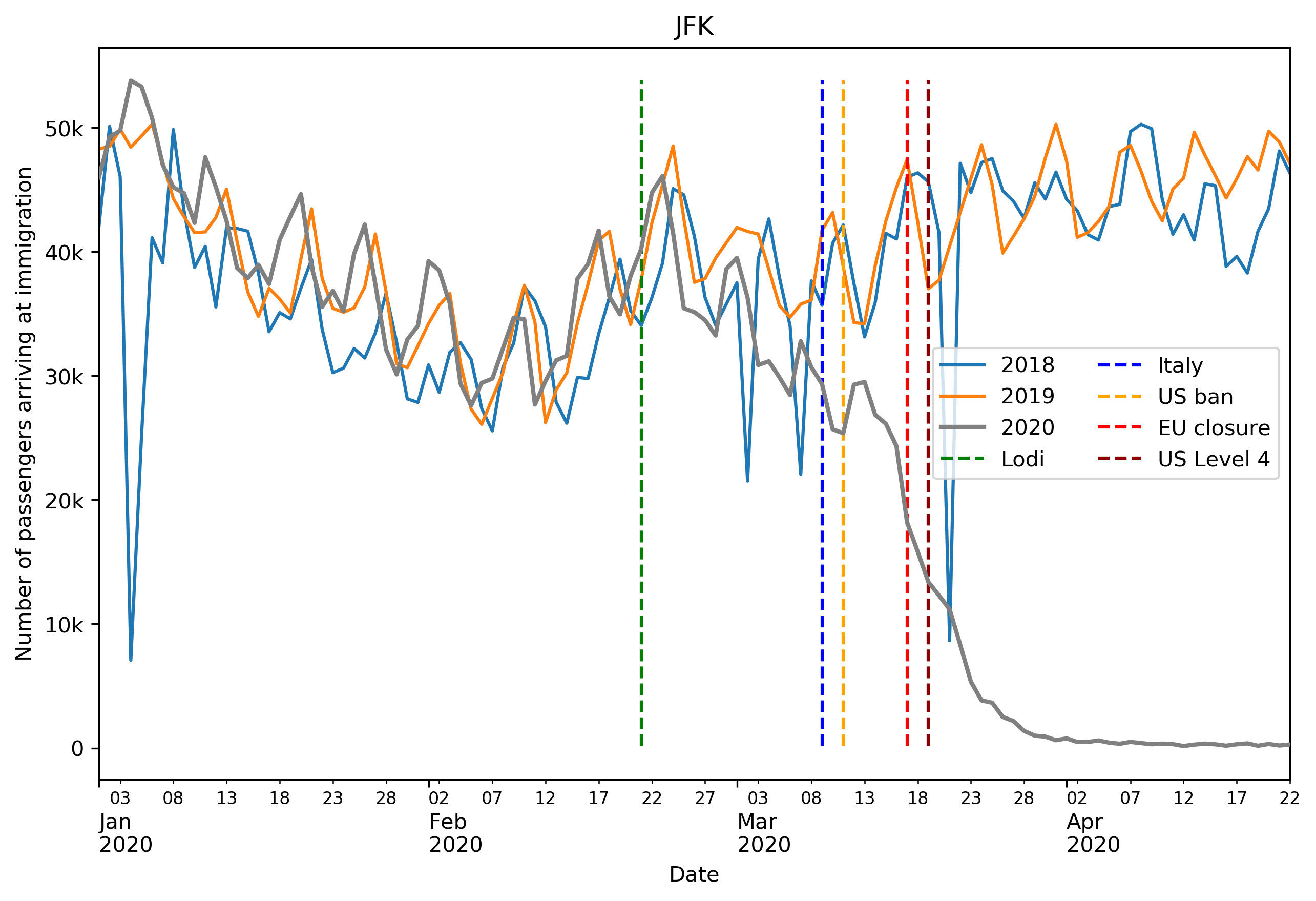}
\caption{Daily number of passengers arriving at immigration}
\label{fig:cbp_cmp_pax_jfk}
\end{subfigure}\hfill
\begin{subfigure}[t]{.49\textwidth}
\includegraphics[width=\textwidth]{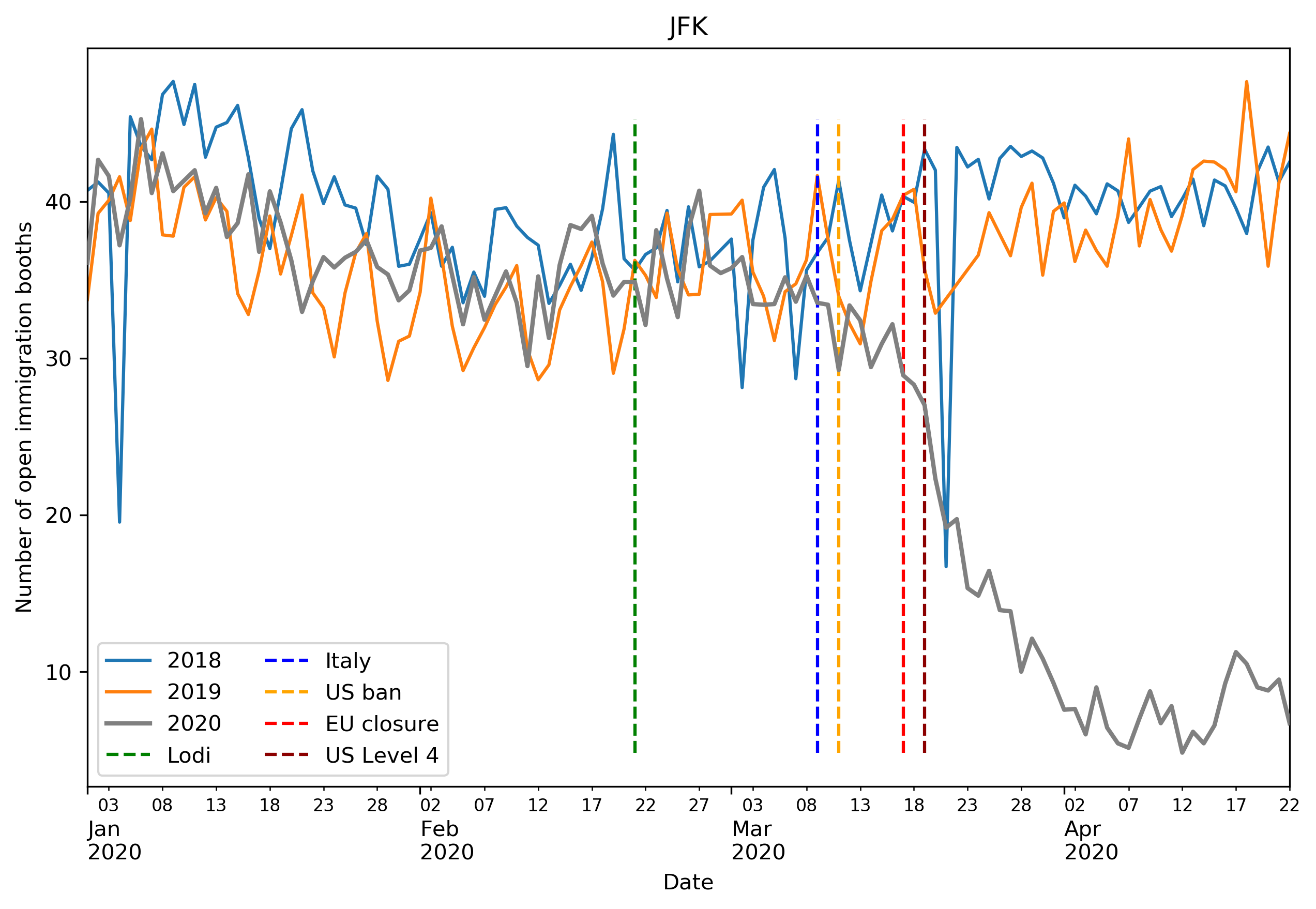}
\caption{Daily average of the number of open immigration booth per hour}
\label{fig:cbp_cmp_booth_jfk}
\end{subfigure}
\begin{subfigure}[t]{.49\textwidth}
\includegraphics[width=\textwidth]{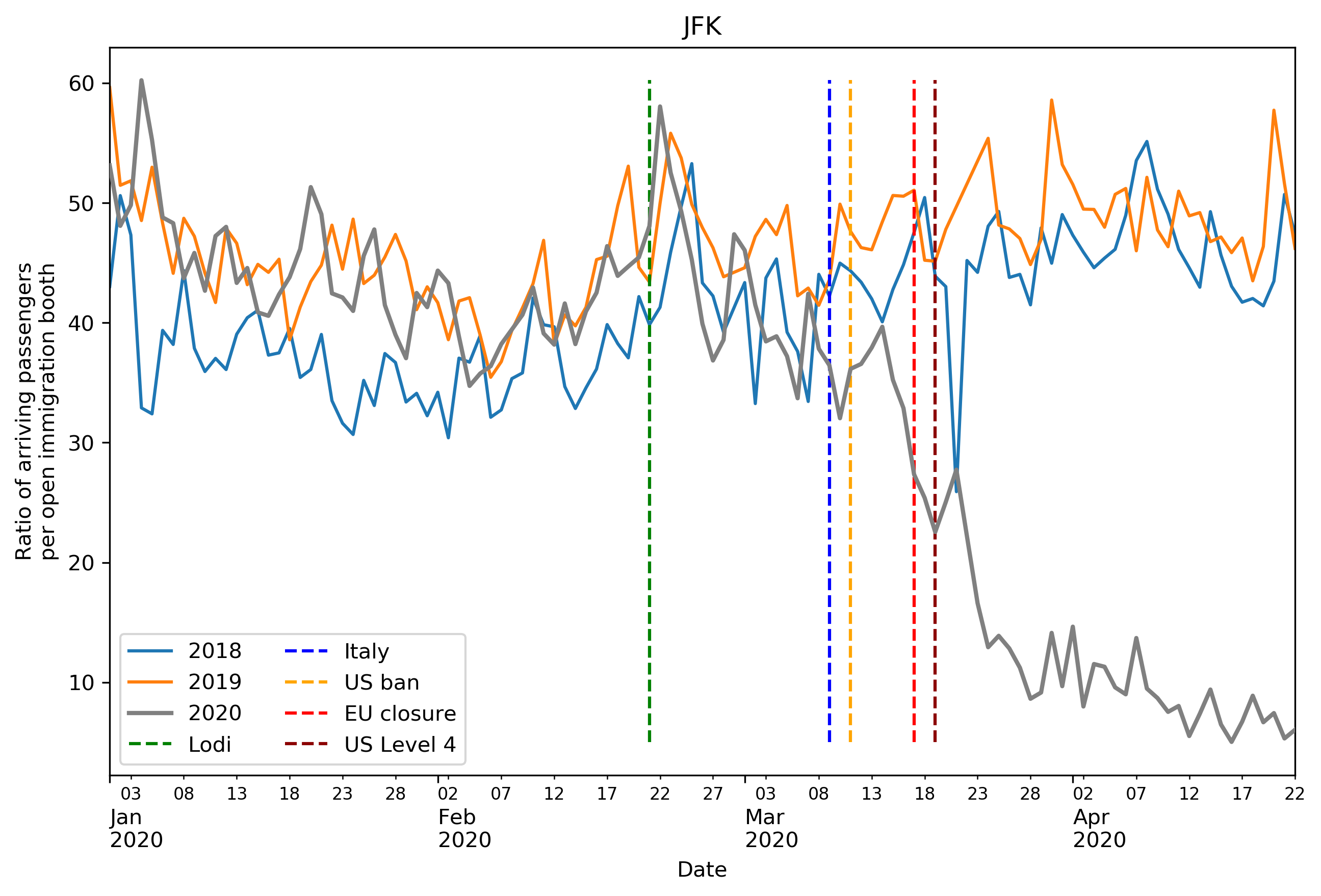}
\caption{Daily average load factor}
\label{fig:cbp_cmp_ratio_jfk}
\end{subfigure}\hfill
\begin{subfigure}[t]{.49\textwidth}
\includegraphics[width=\textwidth]{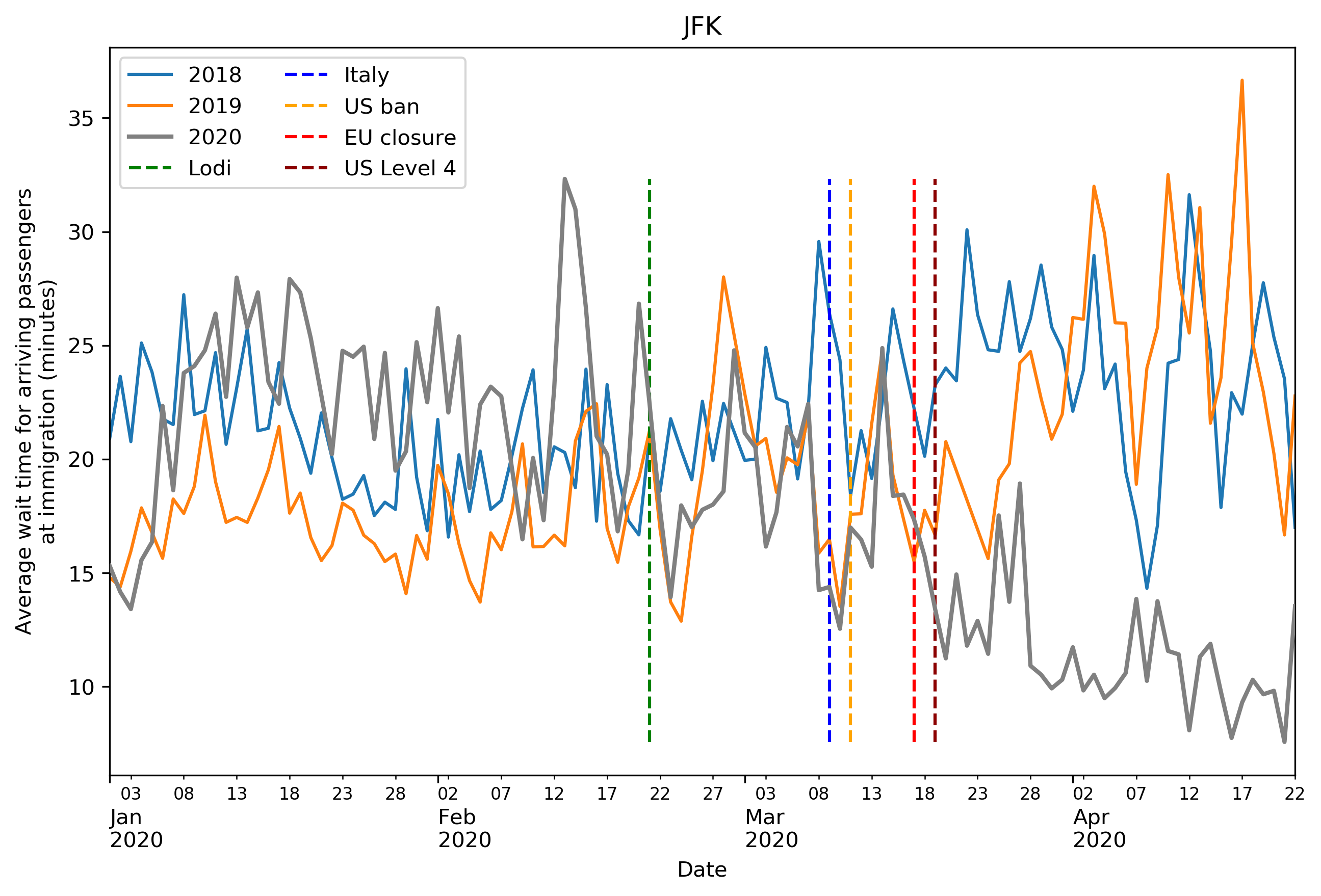}
\caption{Daily average wait time for arriving passengers at immigration}
\label{fig:cbp_cmp_wait_jfk}
\end{subfigure}
\caption{Comparison of CBP data from January 1\st\,to April 13\nth\, for the years 2018 to 2020, case of JFK airport}
\label{fig:cbp_cmp_jfk}
\end{center}
\end{figure}

JFK had the best increase in rank using the proposed immigration quality score presented in Table\,\ref{tab:cbp_ratio_ranking}, and this section aims at analyzing the available CBP immigration data.
The effect of the travel ban measures presented in Section\,\ref{sec:intro_lock} on passengers arriving at JFK's immigration is presented in Figure\,\ref{fig:cbp_cmp_jfk} through four different views by comparing data from 2020 with CBP data from the years 2018 and 2019 between January 1\st\, and April 22\nd.

Figure\,\ref{fig:cbp_cmp_pax_jfk} shows the daily evolution of the number of passengers arriving at JFK's immigration and confirms the important huge drop in the number of arriving international passengers to the US from an average 35.6 thousands passengers arriving at immigration a day down to barely 360 passengers a day. Compared to the years 2018 and 2019, with an average of 45.9 thousands passengers, the difference is more important, since the number of passengers arriving at JFK's immigration is usually higher in April than in March.
Figure\,\ref{fig:cbp_cmp_booth_jfk} shows the daily evolution of the average number of open immigration booths per hour at JFK and presents a similar drop than in Figure\,\ref{fig:cbp_cmp_pax_jfk}, the number of open immigration booths dropping from around an average of 35.4 per operating hour down to an average of 7.5 per hour. % (40 in 2018 and 2019). 
This drop is however less important in proportion compared to the drop in the number of passengers arriving at immigration. Figure\,\ref{fig:cbp_cmp_ratio_jfk} shows the evolution of the daily average load factor (Definition\,\ref{def:load_factor}). After the lockdown and travel ban measures, the load factor drops significantly from an average of 42.5 before the measures down to around 8.5, which represents a 80\% drop. This indicates that after the measures, an immigration booth has about five times fewer passengers to process per hour. Or from the passenger perspective, each passenger has about five times more open immigration booths to take care of them. 
This has a direct positive impact to the average wait time at immigration for passengers.
Figure\,\ref{fig:cbp_cmp_wait_jfk} shows the daily evolution of the average wait time for passengers at JFK's immigration. It was reduced by half after the lockdown and travel ban measures, from around 21.5 minutes to around 10.5 minutes, compared to the usual April levels of 26 minutes in 2019 and 23 minutes in 2018.

\subsubsection{IAD}
\begin{figure}[h!t]
\begin{center}
\begin{subfigure}[t]{.49\textwidth}
\includegraphics[width=\textwidth]{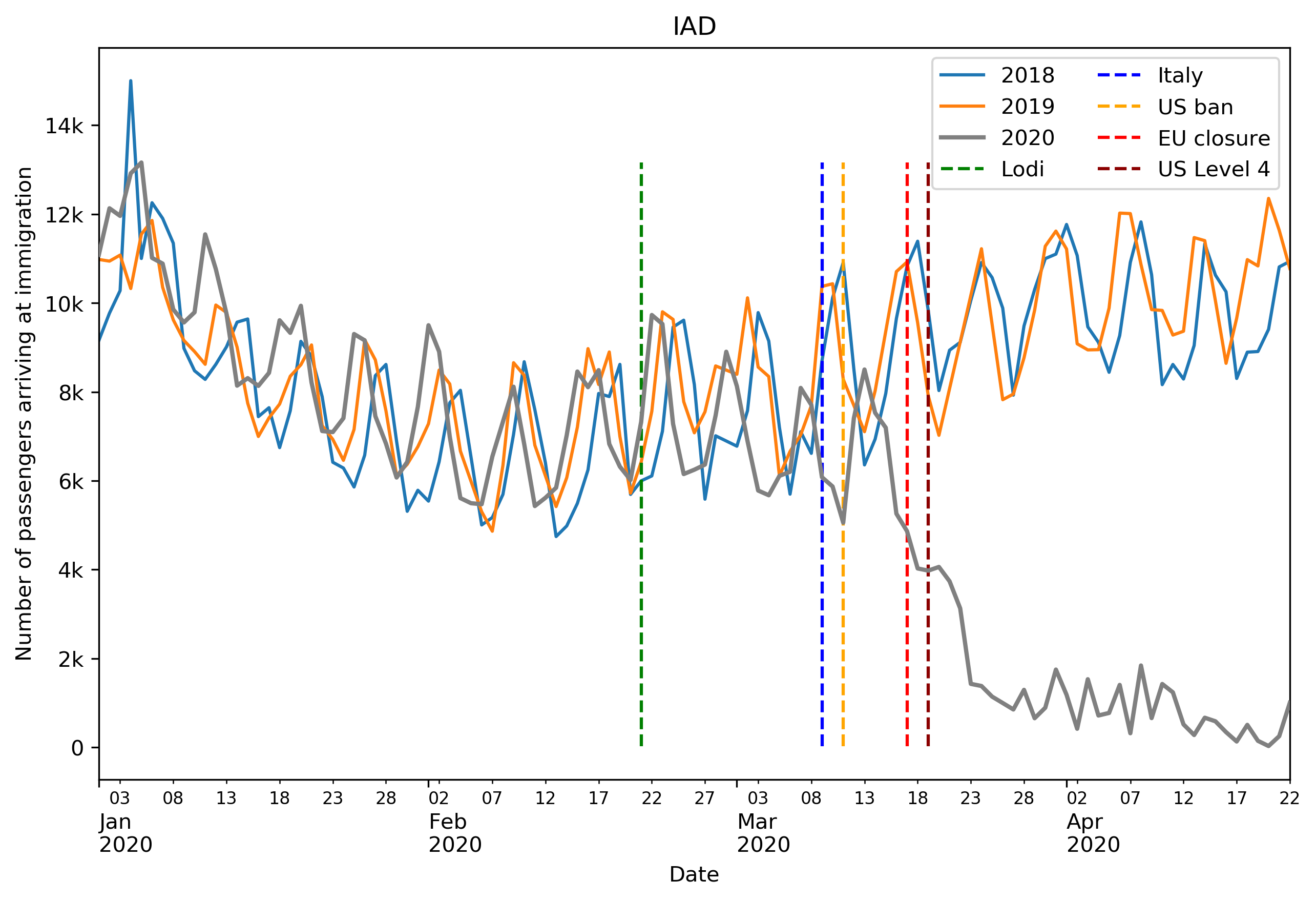}
\caption{Daily number of passengers arriving at immigration}
\label{fig:cbp_cmp_pax_iad}
\end{subfigure}\hfill
\begin{subfigure}[t]{.49\textwidth}
\includegraphics[width=\textwidth]{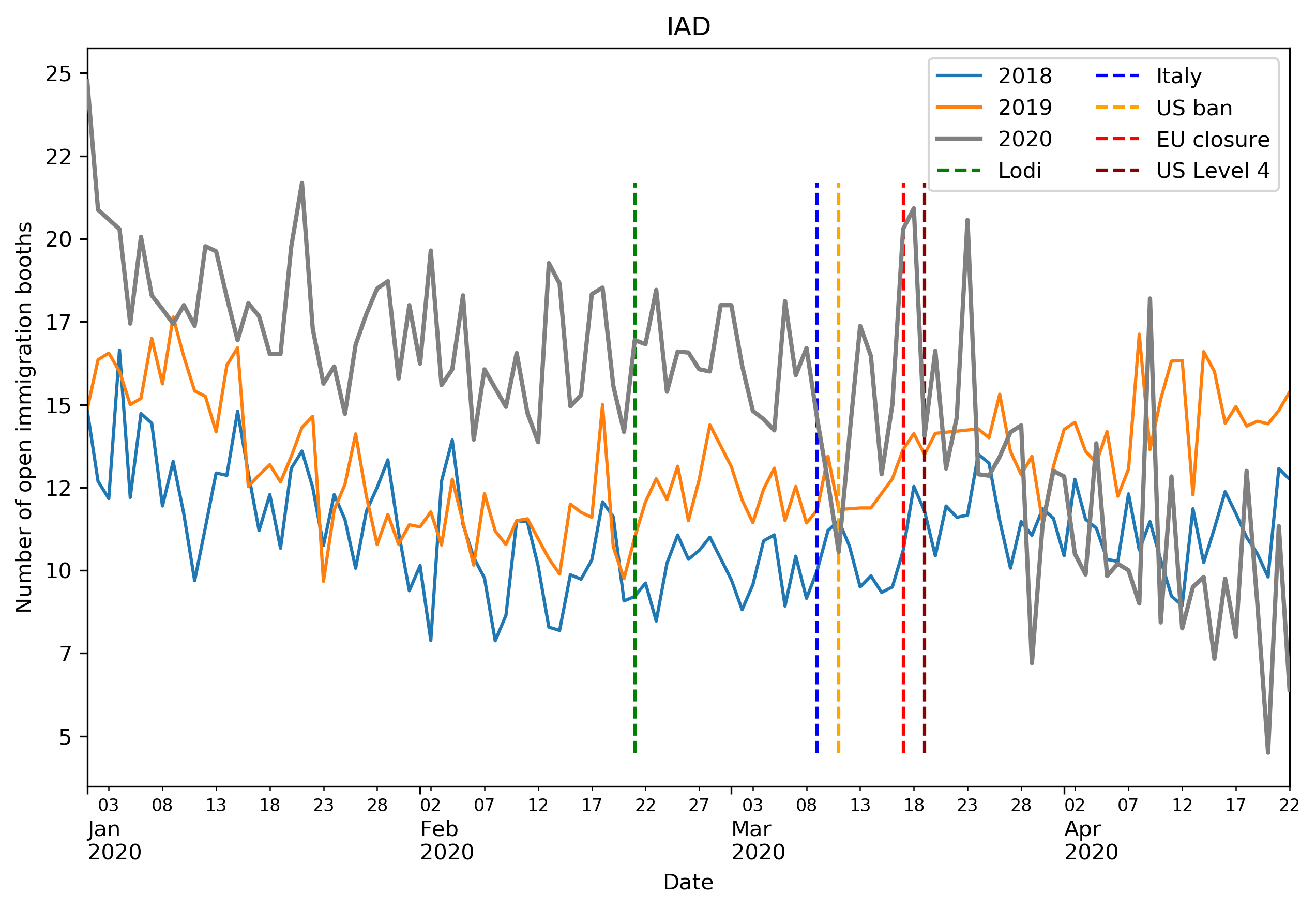}
\caption{Daily average of the number of open immigration booth per hour}
\label{fig:cbp_cmp_booth_iad}
\end{subfigure}
\begin{subfigure}[t]{.49\textwidth}
\includegraphics[width=\textwidth]{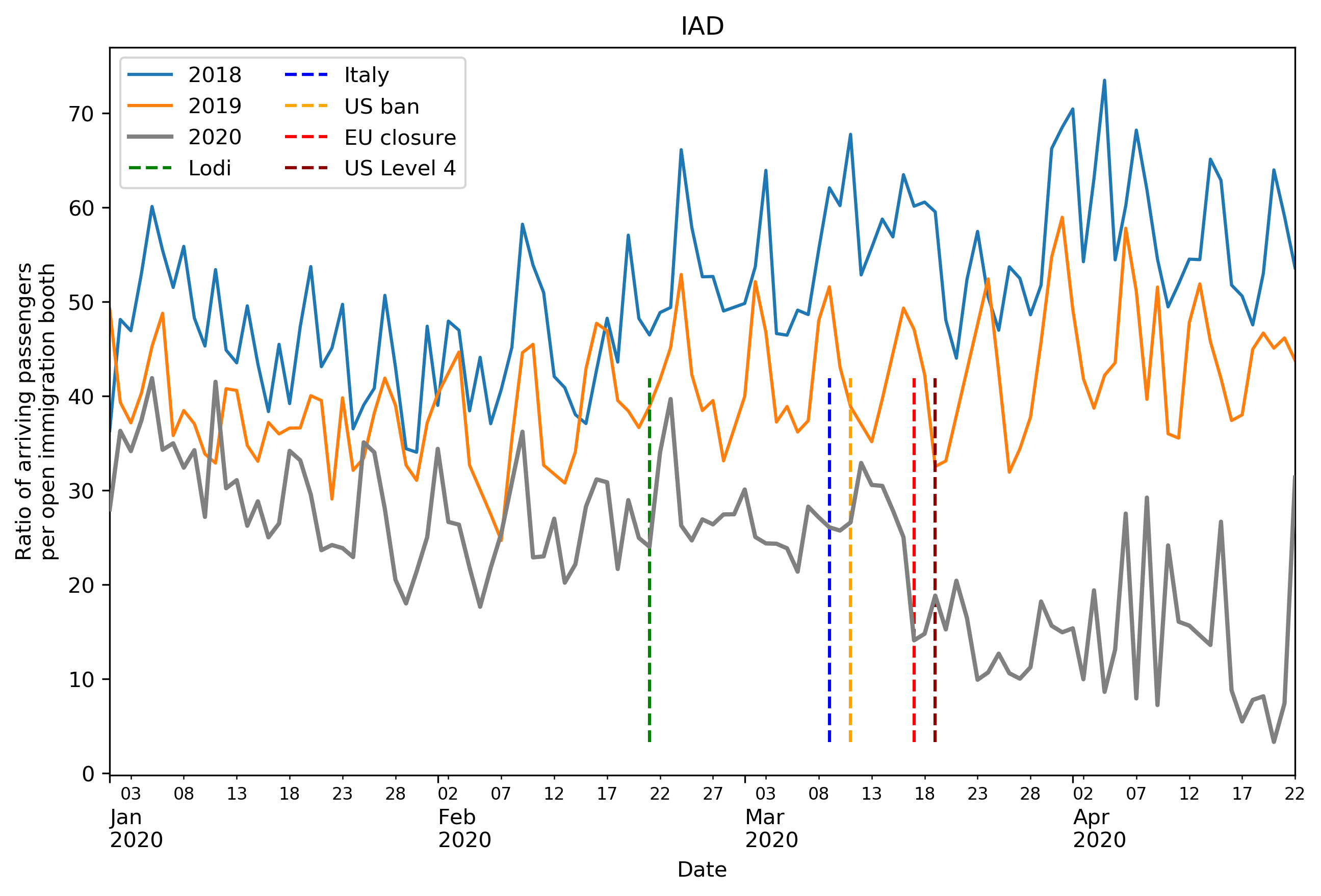}
\caption{Daily average ratio of passengers arriving at immigration per open immigration booths}
\label{fig:cbp_cmp_ratio_iad}
\end{subfigure}\hfill
\begin{subfigure}[t]{.49\textwidth}
\includegraphics[width=\textwidth]{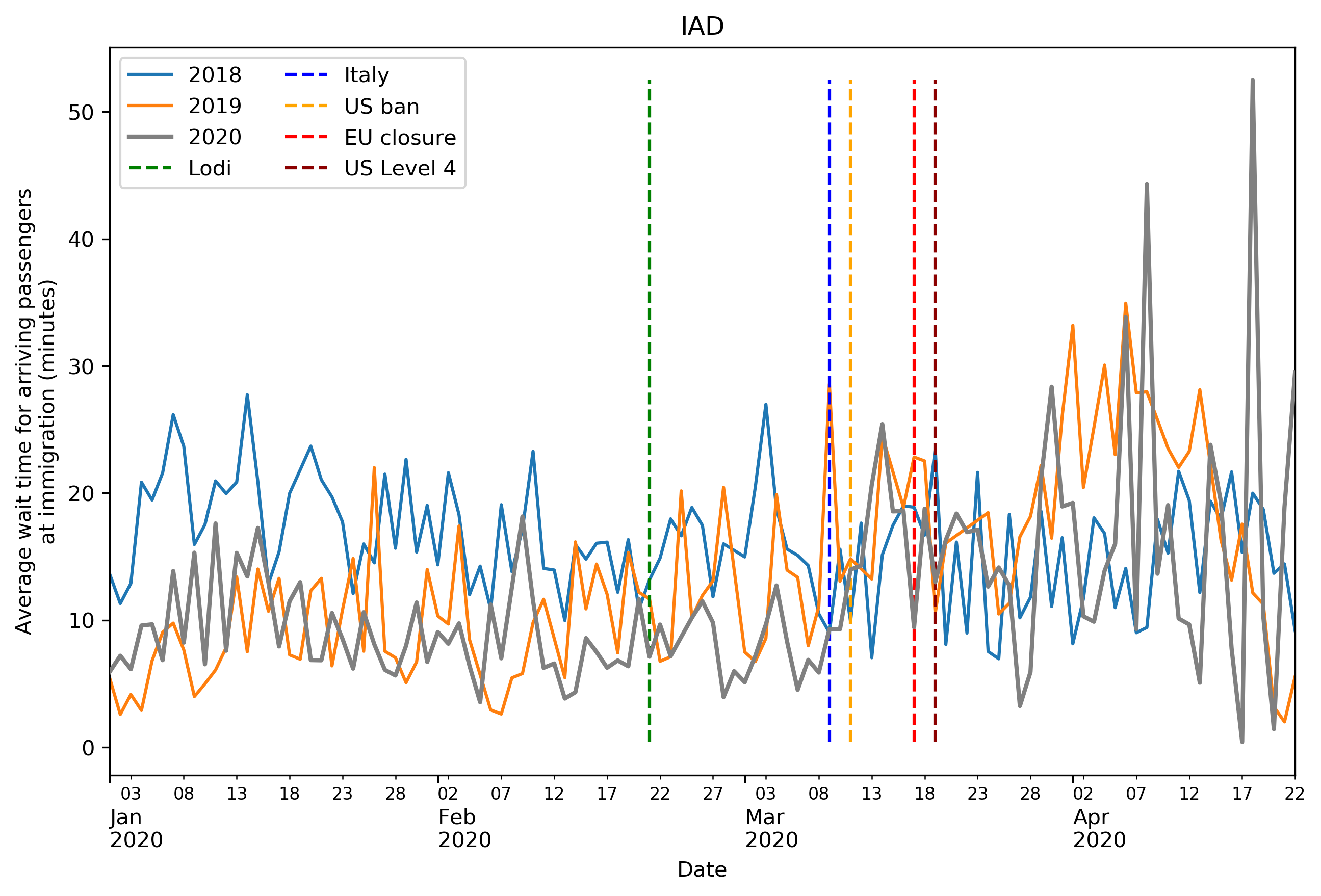}
\caption{Daily average wait time for arriving passengers at immigration}
\label{fig:cbp_cmp_wait_iad}
\end{subfigure}
\caption{Comparison of CBP data from January 1\st\,to April 13\nth\, for the years 2018 to 2020, case of IAD airport}
\label{fig:cbp_cmp_iad}
\end{center}
\end{figure}

IAD had the worst drop in rank using the proposed immigration quality score presented in Table\,\ref{tab:cbp_ratio_ranking}, and is the focus of this section.
Figure\,\ref{fig:cbp_cmp_iad} shows the impact of the travel restriction measures for passengers arriving at IAD's immigration through the four same perspectives as the analysis of JFK.

Figure\,\ref{fig:cbp_cmp_pax_iad} shows the daily evolution of the number of passengers arriving at IAD immigration and confirms that, even though in 2020 that number has dropped from an average of 7.2 thousands in February 2020 to an average of 726 in April 2020 after the implementation of the travel ban measures, the drop is less important than for JFK (Figure\,\ref{fig:cbp_cmp_pax_jfk}).
Though this is still a 93\% drop for the number of passengers arriving at immigration in April between the years 2019 and 2020, with a daily average of 10.4 thousands passengers in 2019, the number of open immigration booths was not impacted as much as for JFK.
Figure\,\ref{fig:cbp_cmp_booth_iad} shows the daily evolution of the average number of open immigration booth per hour at IAD. The daily average of open booths per hour over the month of April 2020, with an average of 10.1 per hour, is similar to the daily average over the month of April 2018, with an average of 11 per hour, and only slightly lower than the number of open booths over the month of April 2019, with an average of 14.6 per hour. Over the period of January to March, the daily average of open booths per hour is significantly higher in 2020 than in 2018 or 2019, with an average of 10.2 for 2018 and of 11.7 for 2019.
From a load factor perspective, the situation at IAD is quite similar to the situation at JFK. Figure\,\ref{fig:cbp_cmp_ratio_iad} shows the daily evolution of the immigration load factor at IAD. The decrease in passengers after the travel ban measures led to a load factor that oscillates around an average of 26.7, which is three to four times lower than the usual load factor of this period. The drop is of 67\% with the year 2019 and of 75\% with the year 2018.
Even though passengers arriving at immigration starting March 20\nth\,2020 have at least three times more available open booths than in the previous year, the wait time for passengers at immigration did not improve, unlike for passengers arriving at JFK immigration.
Figure\,\ref{fig:cbp_cmp_wait_iad} shows the daily evolution of the average wait time for passengers arriving at IAD immigration. The average wait time has increased throughout the travel ban measures and even reached the same level as during the previous years. It went from an average of 8.1 minutes in February 2020 to an average of 17 minutes in April 2020, compared to an average of 14.6 minutes in 2018 and of 26.3 minutes in 2019.
\section{Impact of the COVID-19 related travel restriction measures on US airlines}
\label{sec:airlines}
This section leverages a dataset actively generated by passengers to observe the effects of the travel restriction measures presented in Section\,\ref{sec:intro_lock} on seven major US airlines and propose several passenger-centric metrics to analyze their reactions with respect to their customers.

\subsection{Twitter: a database of passenger-generated content available in real-time}
Airlines operate over more than one airport and in each airport there are usually more than one airline operating at the same time. There is thus no straight-forward way to determine which airline a passenger is flying with using only geolocation data without excessive passenger tracking. The datasets presented and used for the evaluation of the COVID-19 travel restriction measures on airports cannot be directly used to evaluate the impact of these same measures to airlines. Another approach is thus necessary to evaluate the reaction of airlines to this pandemic situation from a passenger perspective.

The importance of airport experience in customer, i.e. passenger, satisfaction towards both airline and airport services was already highlighted in the study of Pruyn and Smidts \cite{pruyn1998EffectsWaitingSatisfaction}, where they show that customer satisfaction is largely affected by their experience at waiting areas, both in terms of wait times and wait environment. This implies that waiting at airports (or any other transit station) can be acceptable for passengers if they are taken care of accordingly. Watkins et al. \cite{watkins2011WhereMyBus} confirmed this conclusion by showing that the perceived wait time for transit riders was lower for riders receiving real-time information than for passengers without that information.

One means of real-time information is social media. In particular, Twitter is an important means of direct communication between airlines and passengers, with an average of more than 300 tweets a day over the month of January 2020 written by the customer services of four major US carriers (Southwest Airlines, Delta Airlines, American Airlines and United Airlines) and an average of more than 800 tweets a day written by their customers. 

The use of Twitter as a real-time estimator of the air transportation system has already been investigated in \cite{monmousseau2019PassengersSocialMediaa} in order to estimate flight-centric values per airport before they were released by BTS, and the same data extraction process is used here. Seven airlines, and their associated Twitter handles, are considered in the following sections: American Airlines (@AmericanAir), Delta Air Lines (@Delta), United Airlines (@united), Alaska Airlines (@AlaskaAir), Southwest Airlines (@SouthwestAir), JetBlue Airways (@JetBlue) and Spirit Airlines (@SpiritAirlines). The first four are legacy airlines, and the last three are low-cost carriers. All tweets written from these airlines Twitter account were scrapped from February 1\st\,2020 to April 12\nth\,2020 and are categorized as "customer service tweets". All tweets written over that same period and mentioning at least one of the airline handles that was not written from the associated airline Twitter account were also scrapped and categorized as "passenger tweets".

\subsection{Sentiment analysis}
\label{sec:sent_analysis}
The same classifiers as in the study conducted in \cite{monmousseau2019PassengersSocialMediaa} are used here to estimate the mood expressed within tweets in order to monitor the real-time evolution of the passenger and airline customer service expressed moods. Each classifier gives a score of 1 if it considers that the tweet expresses a positive sentiment and a score of 0 if it expresses a negative sentiment. In effect, each classifier calculates the probability for a tweet of being positive, and then rounds that probability to the closest integer (0 or 1). The classifiers are here transformed into regressors by considering the probability for a tweet of being classified as positive. The output of all trained regressors are then averaged into one single score, going from 0 for a negative mood to 1 for a positive mood.

\begin{figure}[h!t]
\begin{center}
\begin{subfigure}[b]{.49\textwidth}
\includegraphics[width=\textwidth]{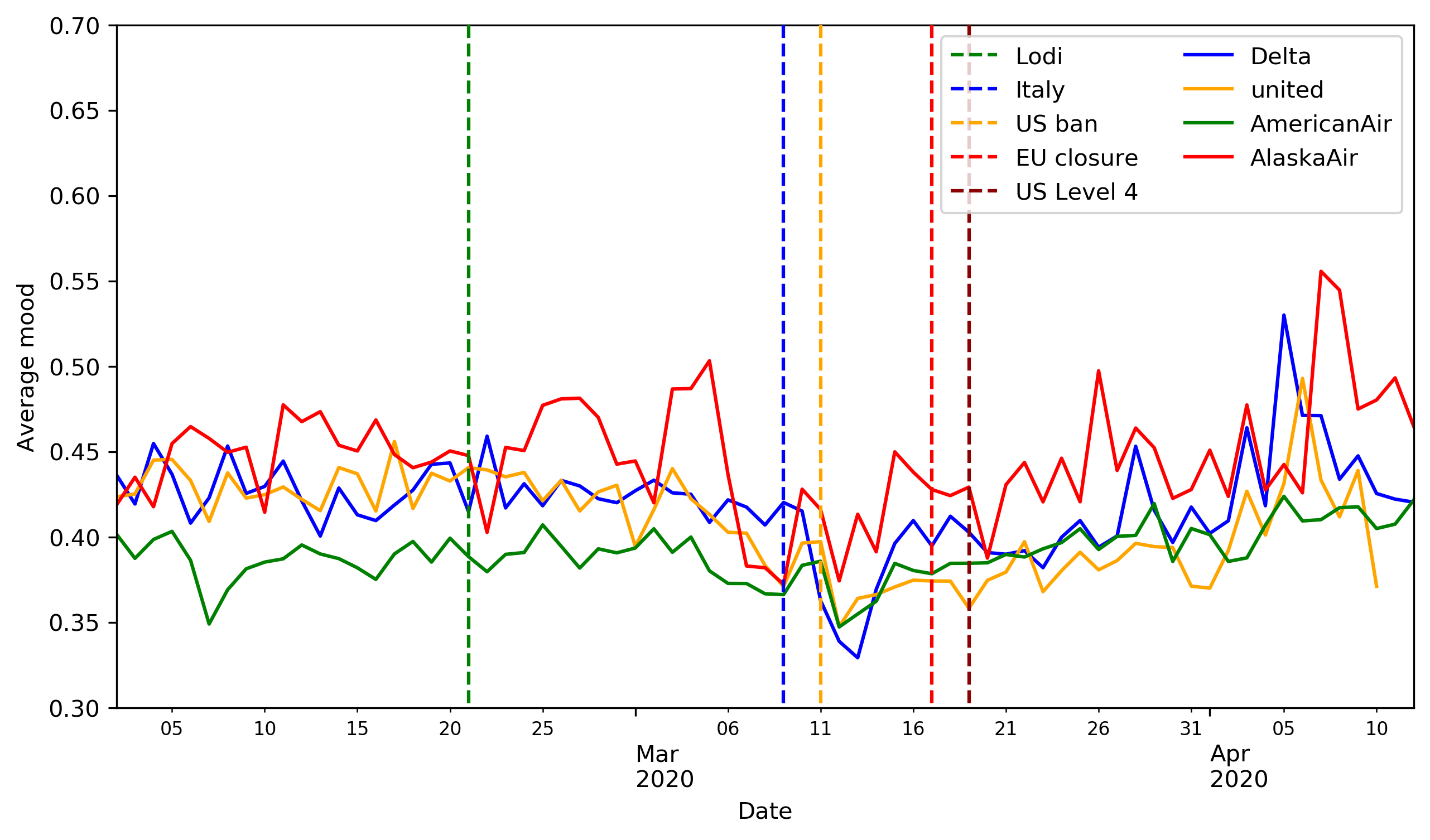}
\caption{From passengers of major airlines}
\label{fig:mean_sent_pax_major}
\end{subfigure}\hfill
\begin{subfigure}[b]{.49\textwidth}
\includegraphics[width=\textwidth]{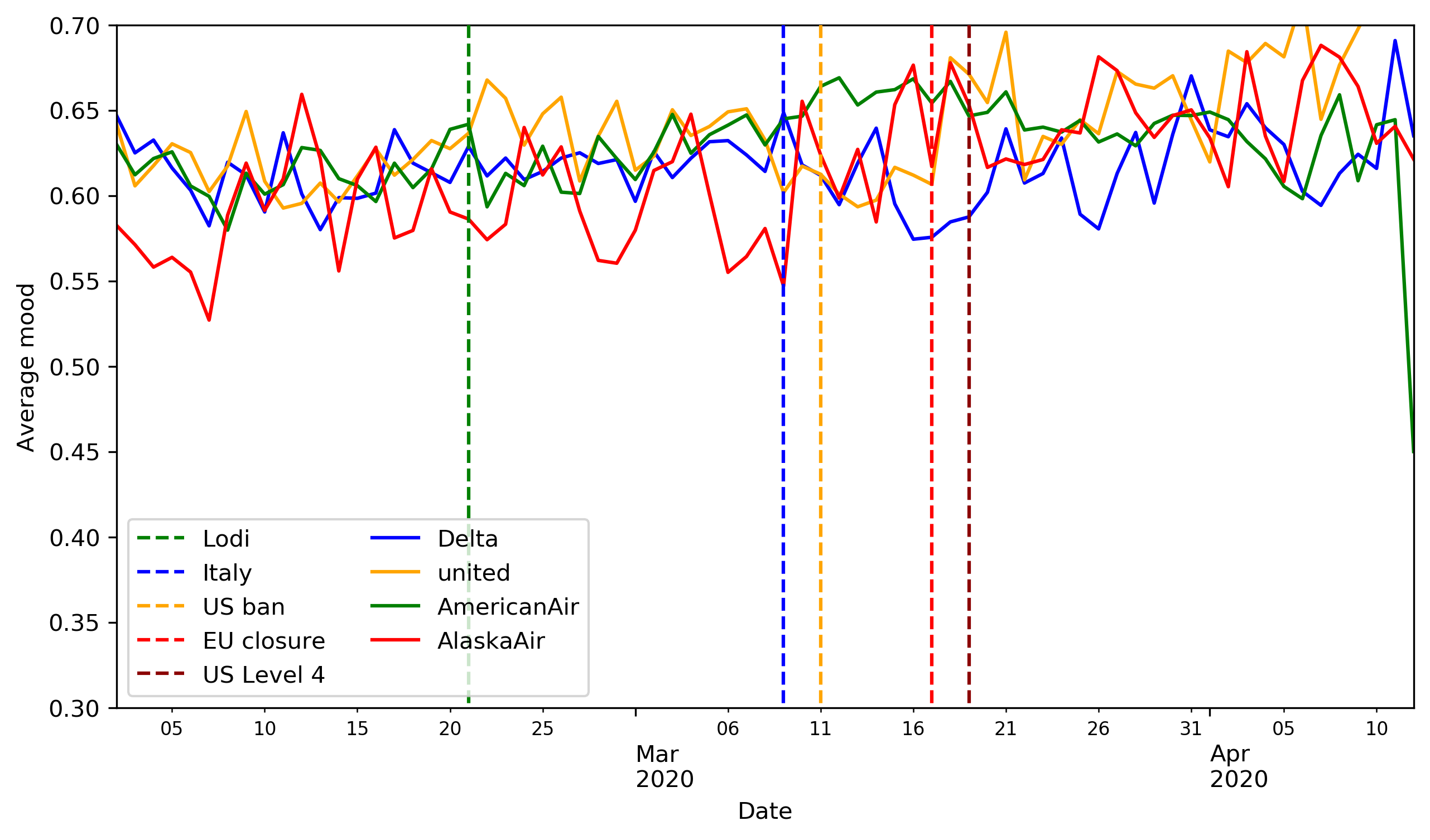}
\caption{From customer service}
\label{fig:mean_sent_cie_major}
\end{subfigure}
\caption{Daily average mood expressed in tweets containing airline Twitter handles for four legacy airlines between February 1\st\,2020 and April 12\nth\, 2020.}
\label{fig:mean_sent_major}
\end{center}
\end{figure}

Using the mean sentiment expressed within each tweets aggregated on a daily level, it is possible to compare the effect of the lockdown from both a passenger perspective and an airline perspective. Figure\,\ref{fig:mean_sent_major} shows the evolution of the expressed mood from February 1\st\,2020 to April 12\nth\, 2020 for the four legacy airlines considered.

From Figure\,\ref{fig:mean_sent_pax_major}, a drop in the mood expressed within passenger tweets can be observed right after the US travel ban for the three major airlines (Delta Air Lines, United Airlines and American Airlines). Delta has the steepest descent but also the sharpest recovery. The case of Alaska Airlines is particular: a \#AlaskaHappyHour campaign, which gave the opportunity of winning free flights to Alaska, took place in the beginning of March 2020. This campaign could explain the increase in the expressed mood in passenger tweets between March 1\st\,2020 and March 5\nth\,2020 and could as well as compensate the effect of the travel ban announcement. 

This drop in the mood is less visible (or non-existent in the case of American Airlines) within the tweets written by the airline customer services, as shown in Figure\,\ref{fig:mean_sent_cie_major}. Though Delta Air Lines and Alaska Airlines had the highest expressed mood on average within passenger tweets, the mood expressed by their customer service is the lowest on average of the four legacy airlines considered. The better mood expressed by their passengers could be explained by the fact that these companies expressed a mood closer to their passengers' actual mood. The gap between the mood expressed within tweets written by passengers and tweets written by airline customer services is visible from one figure to another, airline customer service tweets expressing a mood about 0.2 points higher than passenger tweets.

\begin{figure}[h!t]
\begin{center}
\begin{subfigure}[b]{.49\textwidth}
\includegraphics[width=\textwidth]{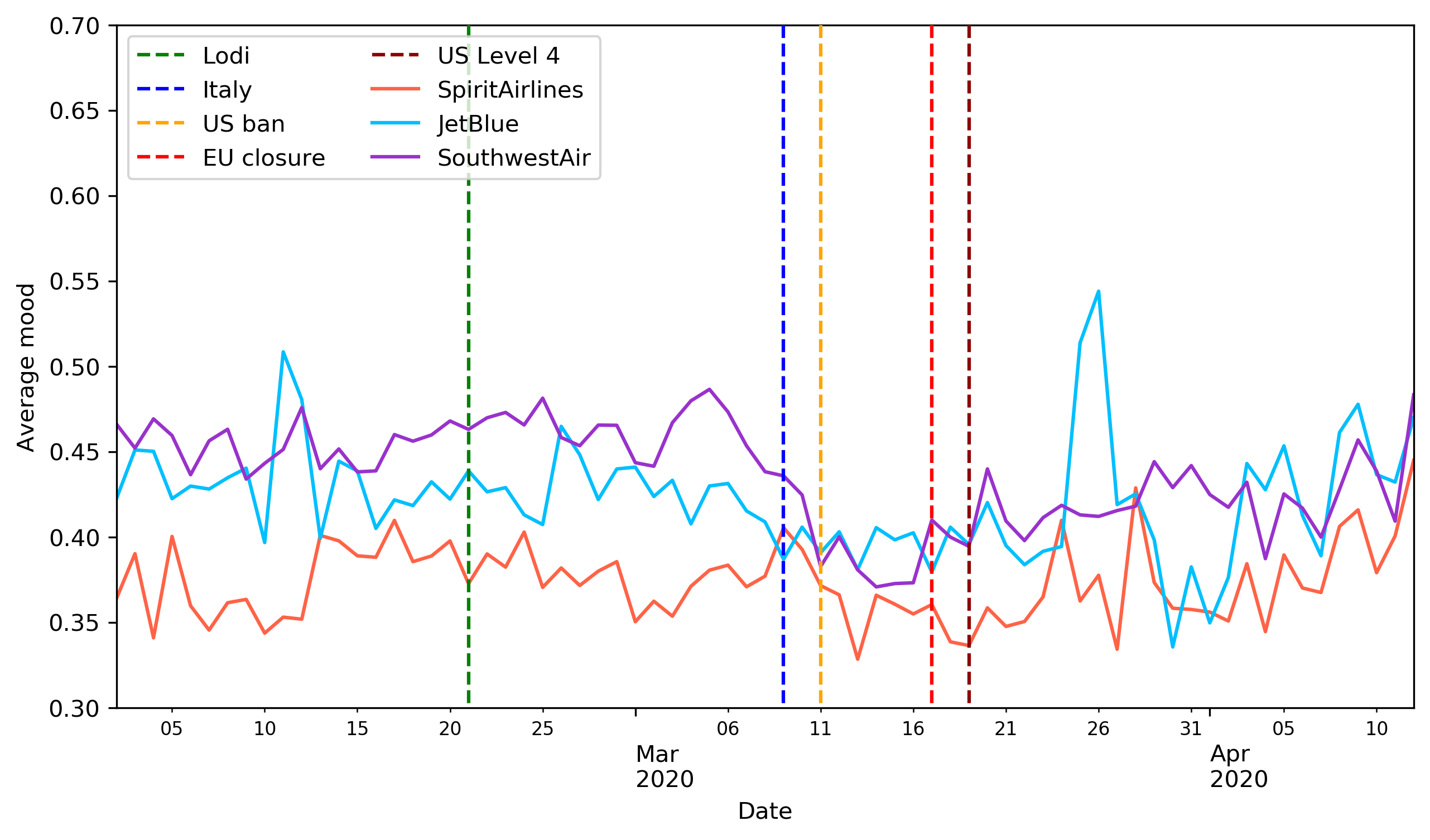}
\caption{From passengers}
\label{fig:mean_sent_pax_lowcost}
\end{subfigure}\hfill
\begin{subfigure}[b]{.49\textwidth}
\includegraphics[width=\textwidth]{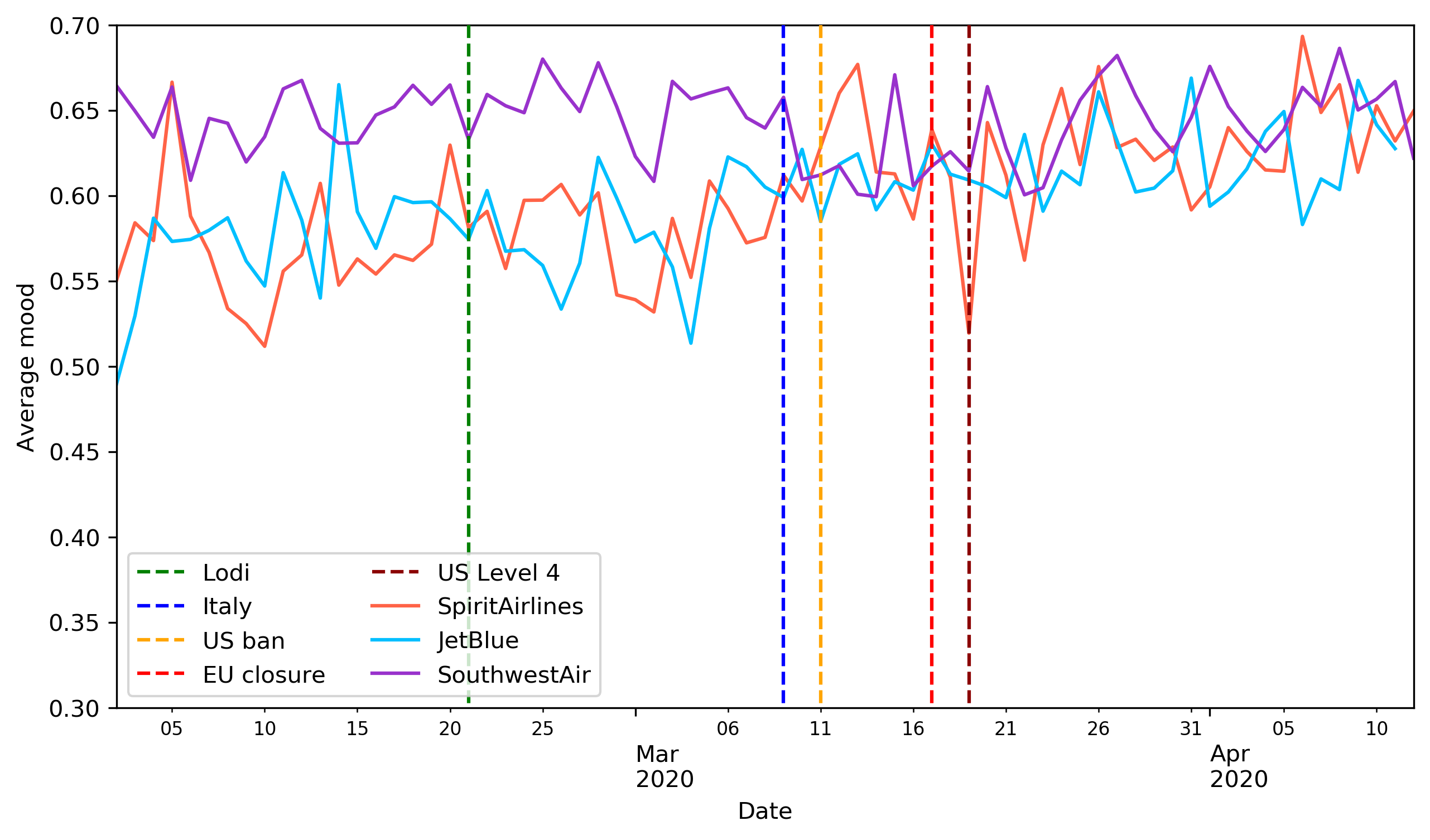}
\caption{From customer service}
\label{fig:mean_sent_cie_lowcost}
\end{subfigure}
\caption{Daily average mood expressed in tweets containing airline Twitter handles for three low-cost airlines between February 1\st\,2020 and April 12\nth\, 2020.}
\label{fig:mean_sent_lowcost}
\end{center}
\end{figure}

Similar observations can be drawn from a low-cost carrier perspective. Figure\,\ref{fig:mean_sent_lowcost} shows the evolution of the expressed mood from February 1\st\,2020 to April 12\nth\, 2020 for the three low-cost carriers considered.

Figure\,\ref{fig:mean_sent_pax_lowcost} indicates that passengers from Spirit Airlines express a significantly lower mood on average than the other two low-cost carriers considered over the months of February and March 2020. A spike in the expressed mood in tweets written by JetBlue passengers can be seen around March 26\nth\,2020, which corresponds to the period when JetBlue announced they would be offering free flights to health care workers in order to help the governor of New York handle the spread of COVID-19 in New York State\footnote{\url{https://twitter.com/NYGovCuomo/status/1242941085535608835}}, as well as the period when an update of their mobile application contained the message "Now, go wash your hands". The drop in the mood expressed in the tweets written by passengers of legacy airlines after Italy's lockdown is less visible for passengers of low-cost airlines.

From a customer service perspective, Figure\,\ref{fig:mean_sent_pax_lowcost}, the gap between the mood expressed in the tweets written by Southwest Airlines customer and the mood expressed in the tweets written by the customer service of the other two carriers is resorbed the day after Italy's lockdown. This could indicate a similar communication policy for these carrier regarding the COVID-19 pandemic. The same gap as for legacy airlines between the mood expressed within tweets written by passengers and tweets written by airline customer services of about 0.2 points is visible from Figure\,\ref{fig:mean_sent_pax_lowcost} to Figure\,\ref{fig:mean_sent_cie_lowcost}.

Based on this sentiment data, two metrics are proposed here to compare airlines.
A first metric aims to measure how well airlines are in phase with the mood of their passengers. For example, if the average passenger mood is decreasing, the mood expressed by the airline customer service should not be increasing.
\begin{paxmetric}
\label{pax:empathy}
The airline \textbf{empathy} score is defined as the correlation between the evolution of the average mood expressed by passengers in their tweets and the evolution of the average mood expressed by the airline customer service in their tweets.
\end{paxmetric}
This score goes from -1 to 1, with 1 meaning that the airline customer service expressed mood is perfectly in phase with the mood expressed by their passengers. A score of 0 indicates that the mood expressed by the airline customer service  is uncorrelated with the expressed mood of their passengers. A score of -1 indicates that the airline customer service expressed mood is in complete opposition of phase with the mood expressed by their passengers. In other words, the mood expressed by airlines increases when the mood expressed by passengers decreases, and \emph{vice-versa}.

\begin{paxmetric}
\label{pax:sent_gap}
The airline \textbf{sentiment gap} measures the average difference between the average mood expressed by passengers and the average mood expressed by airlines.
\end{paxmetric}
This measures goes from -1 to 1 with 0 indicating that airlines and passengers express the same average mood. 1 indicates the worst case, i.e. when the mood expressed by airlines is equal to 1 (i.e. highest possible) and the mood expressed by passengers is equal to 0 (lowest possible) throughout the considered period. A measure of -1 indicates the opposite scenario.

Table\,\ref{tab:sent_scores} presents these two proposed passenger-centric metrics based on the sentiment expressed in tweets for the seven considered airlines and ranks the airlines for each score. The scores were calculated over the period of March 1\st\,2020 to April 11\nth\,2020.

\begin{table}[ht]
\begin{center}
\caption{Airline ranking based on the proposed passenger-centric metrics based on the sentiment expressed in tweets applied to the period of March 1\st\,2020 to April 11\nth\,2020}
\label{tab:sent_scores}
\begin{tabular}{|c||c|c||c|c|}
\hline
\bfseries Rank & \bfseries Airline & \bfseries Empathy & \bfseries Airline & \bfseries Sentiment Gap \\ \hline
1&AlaskaAir&0.57&AlaskaAir&0.19\\
2&united&0.45&JetBlue&0.20\\
3&SouthwestAir&0.44&Delta&0.20\\
4&SpiritAirlines&0.23&SouthwestAir&0.22\\
5&JetBlue&0.22&SpiritAirlines&0.25\\
6&Delta&0.17&AmericanAir&0.25\\
7&AmericanAir&-0.52&united&0.25\\

\hline
\end{tabular}
\end{center}
\end{table}

\subsection{Keyword-based metrics}
\subsubsection{Cancellations}
\label{sec:kw_cancel}
When a some exceptional situation occurs, a spike in the use of certain keywords can be seen within the stream of tweets written by the affected Twitter users. For example, in the case of an important number of cancellations, many passengers will go on Twitter and use the keyword "cancel" to express their concerns directly to the airline they were flying with.

\begin{figure}[ht]
\begin{center}
\begin{subfigure}[b]{.49\textwidth}
\includegraphics[width=\textwidth]{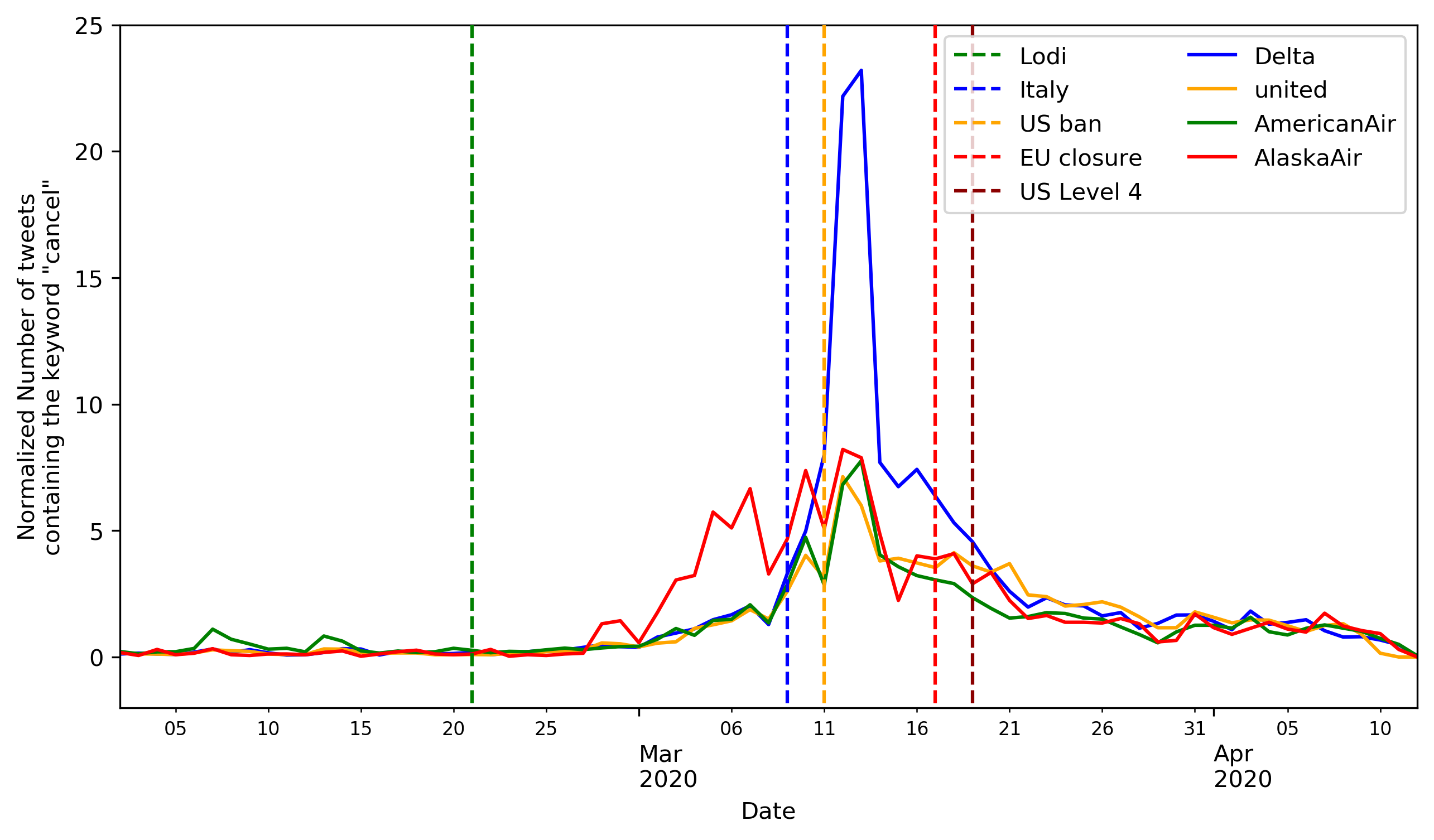}
\caption{From passengers of legacy airlines}
\label{fig:kw_cancel_pax_major}
\end{subfigure}\hfill
\begin{subfigure}[b]{.49\textwidth}
\includegraphics[width=\textwidth]{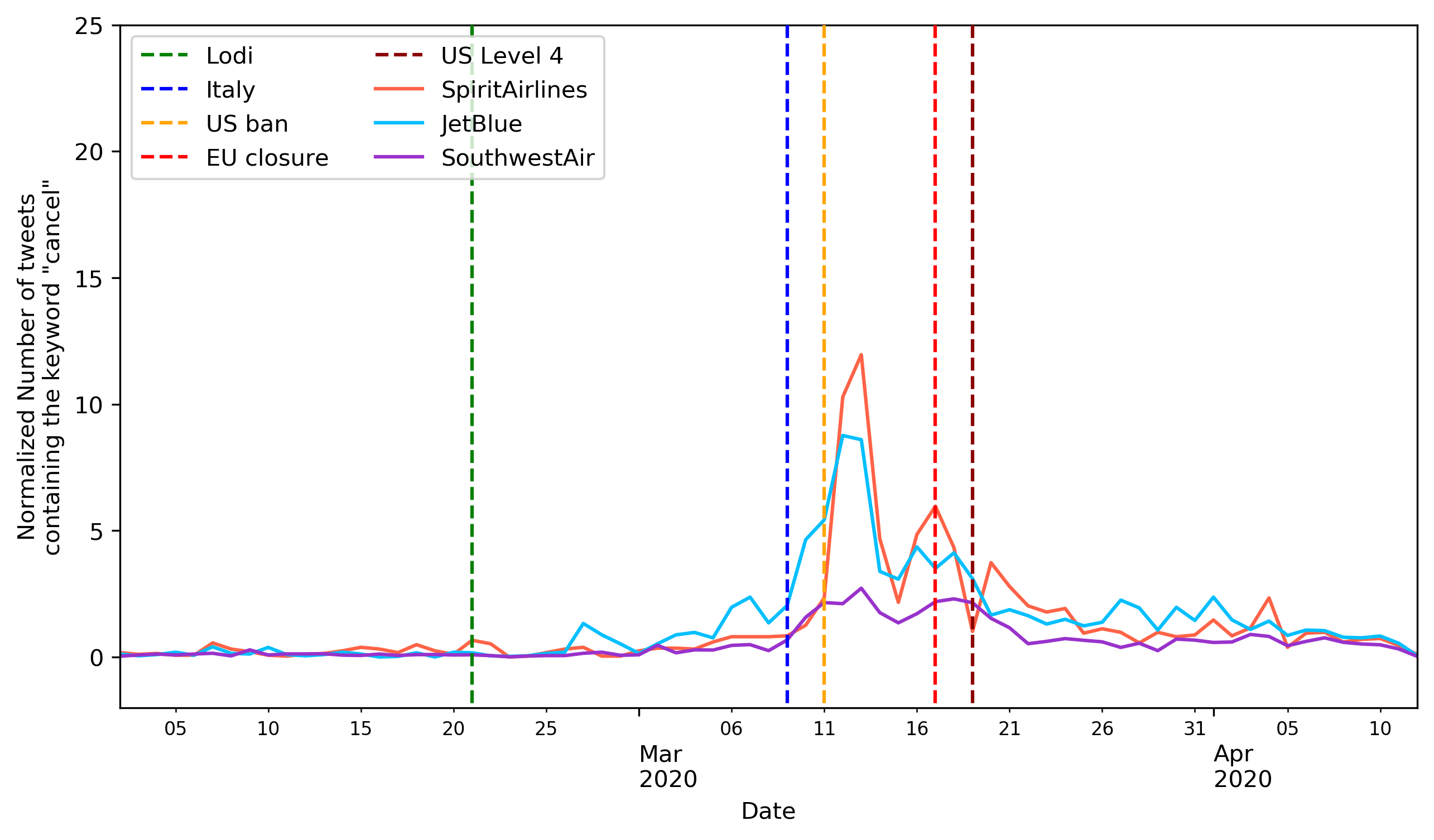}
\caption{From passengers of low-cost airlines}
\label{fig:kw_cancel_pax_lowcost}
\end{subfigure}
\caption{Number of occurrences of the keyword "cancel" in tweets written by passengers normalized by the number of total passengers per carrier over the year 2018 using BTS data \cite{BTS2018carrier}}
\label{fig:kw_cancel_pax}
\end{center}
\end{figure}

Figure\,\ref{fig:kw_cancel_pax} shows the evolution of the normalized number of occurrences of the keyword "cancel" in tweets written by passengers from February 1\st\,2020 to April 12\nth\, 2020 for four US legacy airlines and three US low-cost carriers. The normalization is based on the total number of passengers transported by each considered carrier over the year 2018 using BTS data \cite{BTS2018carrier}.

Figure\,\ref{fig:kw_cancel_pax_major} indicates that the passengers of the four legacy airlines considered are reactive to the international situation; an important increase can already be seen around the date of Italy's lockdown announcement. A second spike then occurs once the US announces that it bans all travellers from the EU, China and Iran, with Delta Air Line passengers being, in proportion, the most vocal on Twitter. This could indicate that Delta Air Line has a greater proportion of its US passengers traveling in the EU at that time. Alaska Airlines had an early spike in the number of tweets containing the keyword "cancel" compared to the other legacy airlines. That early spike could be link to the fact that most of the first US cases of COVID-19 were discovered on the US West Coast, which is where Alaska Airlines main hub is located. 

Figure\,\ref{fig:kw_cancel_pax_lowcost} shows the evolution in the mood expressed in passenger tweets for the three considered low-cost carriers. Passengers of Southwest Airlines are the less vocal in proportion on the matter of cancellation, with a slight increase in number of occurrences of the keyword "cancel" in their tweets almost entirely contained within the period between Italy lockdown announcement and the rise to a Level y travel advisory for the US. Passengers of JetBlue Airways have a behavior similar to legacy airlines in this case. Spirit Airlines passengers waited for the US travel ban announcement to express massively their concerns using the word "cancel".

\begin{figure}[ht]
\begin{center}
\begin{subfigure}[b]{.49\textwidth}
\includegraphics[width=\textwidth]{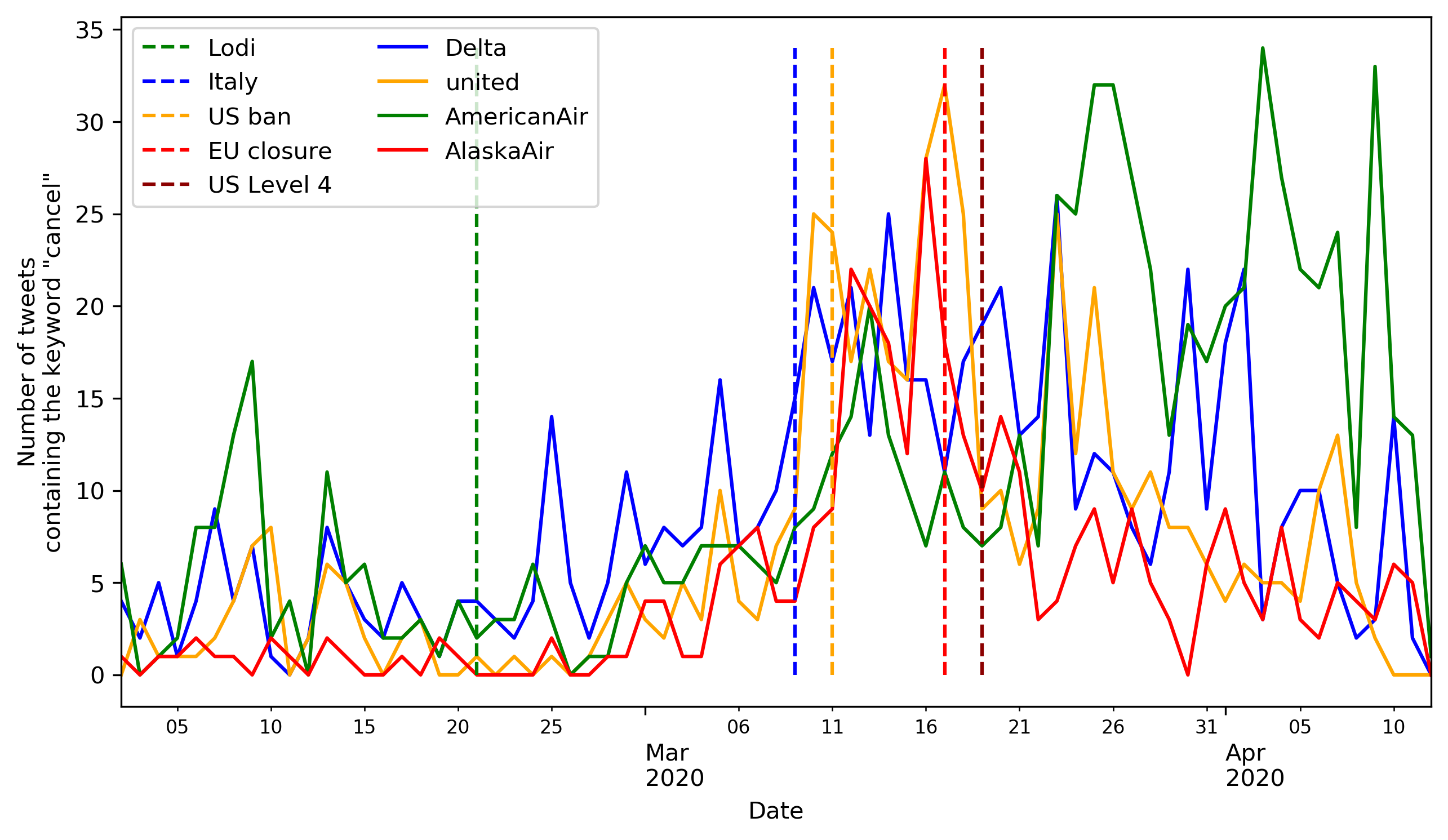}
\caption{From customer service of legacy airlines}
\label{fig:kw_cancel_cie_major}
\end{subfigure}\hfill
\begin{subfigure}[b]{.49\textwidth}
\includegraphics[width=\textwidth]{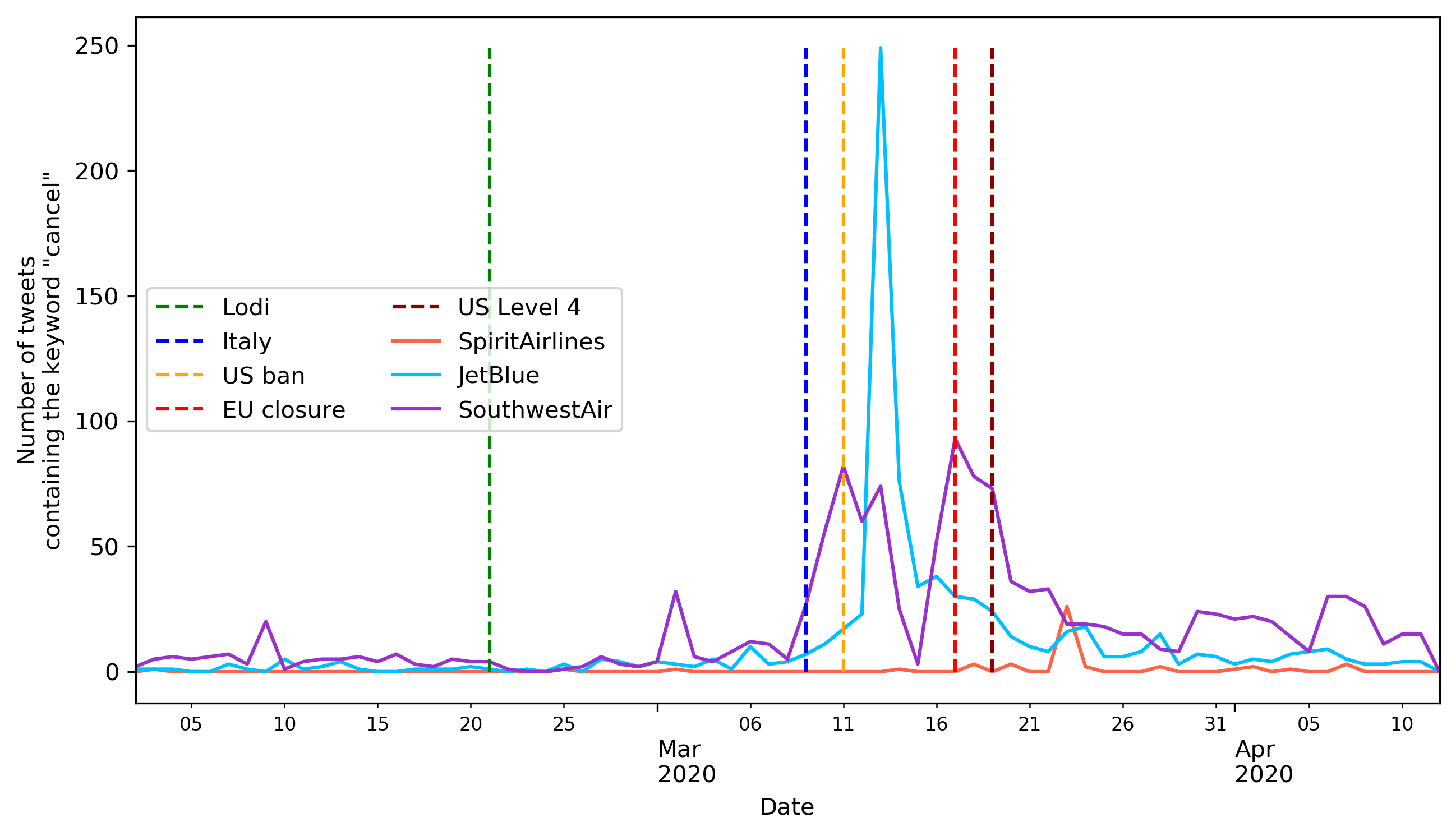}
\caption{From customer service of low-cost airlines}
\label{fig:kw_cancel_cie_lowcost}
\end{subfigure}
\caption{Number of occurrences of the keyword "cancel" in tweets written by airline customer services}
\label{fig:kw_cancel_cie}
\end{center}
\end{figure}

Figure\,\ref{fig:kw_cancel_cie} shows the evolution of the number of occurrences of the keyword "cancel" in tweets written by airline customer sservices from February 1\st\,2020 to April 12\nth\, 2020 for the same four US legacy airlines and three US low-cost carriers. Please note that the y-axis scale is different between Figure\,\ref{fig:kw_cancel_cie_major} and Figure\,\ref{fig:kw_cancel_cie_lowcost}.

For legacy airlines, the behavior shown in Figure\,\ref{fig:kw_cancel_cie_major} is similar for three out of four of the considered airlines. There is a significant increase in the number of occurrences of the keyword "cancel" starting the Italy announced its lockdown and that number then slowly decreases. For American Airlines, after a similar increase in the number of occurrences of the keyword "cancel", that number does not decrease but fluctuates at a level more important than during the period before the travel restriction measures where announced.

Regarding low-cost carriers, Figure\,\ref{fig:kw_cancel_cie_lowcost} shows that each carrier has its own characteristic regarding the use of the keyword "cancel". Southwest Airlines has two important spikes around each of the US announcements referenced in the plot. JetBlue has a single massive spike on March 13\nth\,2020. Both of these carriers then spent more than two weeks with a higher level of occurrences of the keyword "cancel" than in February 2020. Spirit Airlines barely uses the keyword "cancel" in their communication except on March 23\rd\,2020.

Based on the observations from the plots in Figure\,\ref{fig:kw_cancel_pax}, it is possible to consider that an important increase in the normalized number of passenger tweets containing the keyword "cancel" represents a situation that airlines have to deal with in order for that volume to return to normal values.
\begin{mydef}
\label{def:kw_situation}
A keyword-related \textbf{Twitter situation} is defined as an increase in the normalized number of occurrences of the keyword within passenger tweets over a predefined level.
\end{mydef}

Two metrics to measure the airline reaction to such a situation are proposed here.
The aim of the first metric is to measure the effectiveness of the airline response to these keyword-related situations.
\begin{paxmetric}
\label{pax:kw_quality}
The keyword-related Twitter situation \textbf{quality response} score of an airline is the time needed for the airline to bring the normalized number of occurrences of the keyword within passenger tweets back to a predefined level.
\end{paxmetric}
This proposed quality metric measures the time needed for the airline to bring the number of keyword occurrences back to a normal state. When measuring the response of long term perturbation, such as the COVID pandemic, this time is measured in days. 

The number of keyword occurrences in the passenger tweets is normalized by the total number of passengers over the year 2018 in this case, similarly to the data presented in Figure\,\ref{fig:kw_cancel_pax}, and this normalization should be updated with the most recent numbers once they are available. 

The aim of the second metric is to measure the communication effort produced by the airline in order to handle the situation linked to the increase of number of occurrences of the keyword under consideration.
\begin{paxmetric}
\label{pax:kw_quantity}
The keyword-related Twitter situation \textbf{quantity response} score of an airline is calculated by integrating the number of occurrences of the keyword in tweets written by the airline over the number of days associated to the situation.
\end{paxmetric}
The number of days used to calculate this quantity response score corresponds to the number of days found using the quality response score associated with the same situation.

Table\,\ref{tab:kw_cancel_ranking} presents these two proposed metrics in the case of the keyword "cancel" considering that the predefined threshold indicating when a situation starts and ends is 1.
\begin{table}[ht]
\begin{center}
\caption{Airline ranking based on the proposed keyword metrics applied to the period of March 1\st\,2020 to April 11\nth\,2020 with the keyword "cancel"}
\label{tab:kw_cancel_ranking}
\begin{tabular}{|c||c|c||c|c|}
\hline
\bfseries Rank & \bfseries Airline & \bfseries Quality & \bfseries Airline & \bfseries Quantity \\ \hline
1&SouthwestAir&12&SouthwestAir&55.33\\
2&SpiritAirlines&19&JetBlue&22.03\\
3&AmericanAir&32&AmericanAir&15.53\\
4&JetBlue&32&Delta&13.66\\
5&Delta&35&united&12.19\\
6&AlaskaAir&35&AlaskaAir&8.60\\
7&united&36&SpiritAirlines&1.95\\

\hline
\end{tabular}
\end{center}
\end{table}
Table\,\ref{tab:kw_cancel_ranking} illustrates the necessity of considering both the quality response score and the quantity response score hand in hand. Southwest Airlines has the best scores from both perspective but Spirit Airlines has the second best quality response score but the worst quantity response score by far.

\subsubsection{Refund}
\label{sec:kw_refund}
\begin{figure}[h!t]
\begin{center}
\begin{subfigure}[b]{.49\textwidth}
\includegraphics[width=\textwidth]{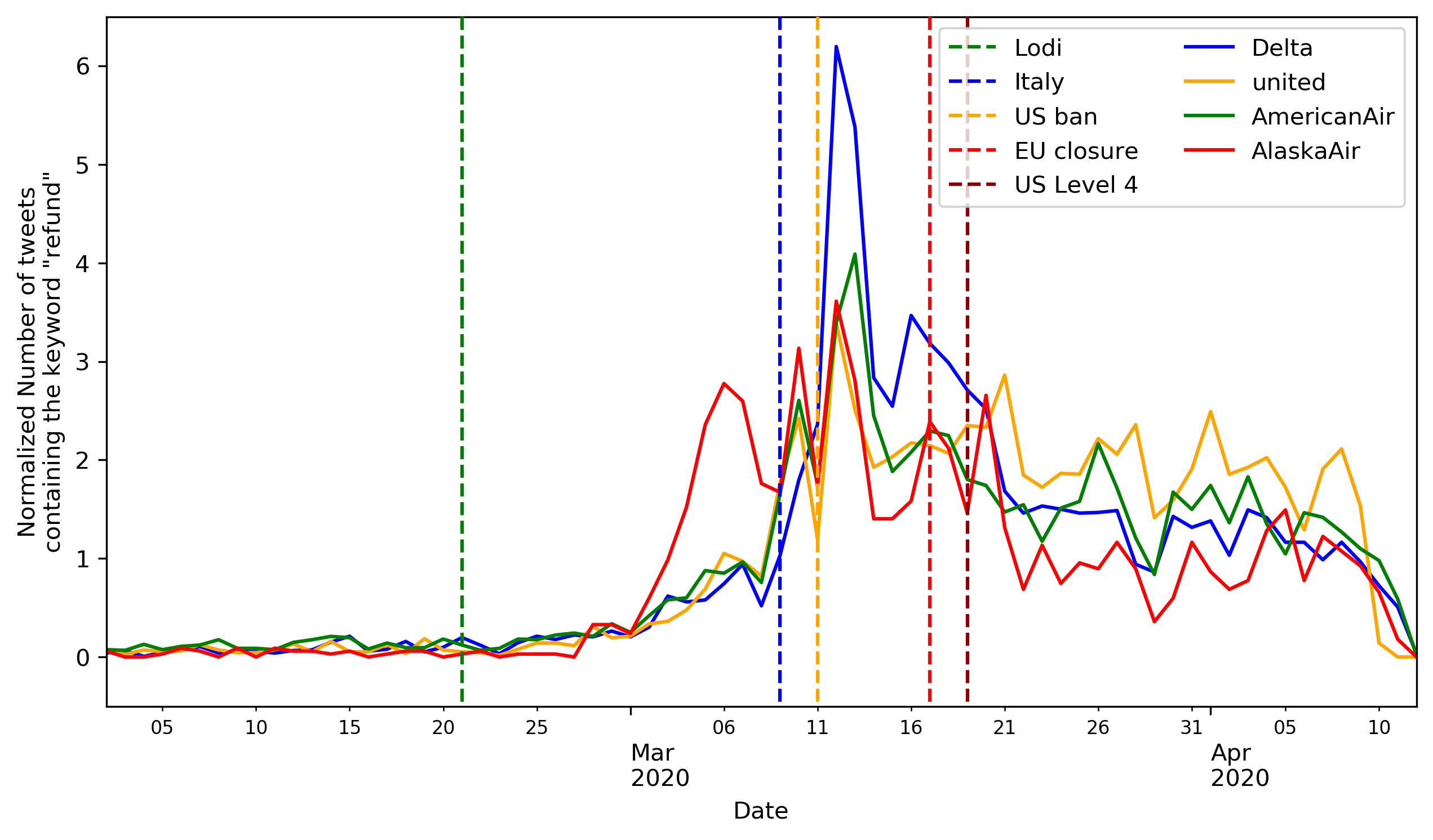}
\caption{From passengers of legacy airlines}
\label{fig:kw_refund_pax_major}
\end{subfigure}\hfill
\begin{subfigure}[b]{.49\textwidth}
\includegraphics[width=\textwidth]{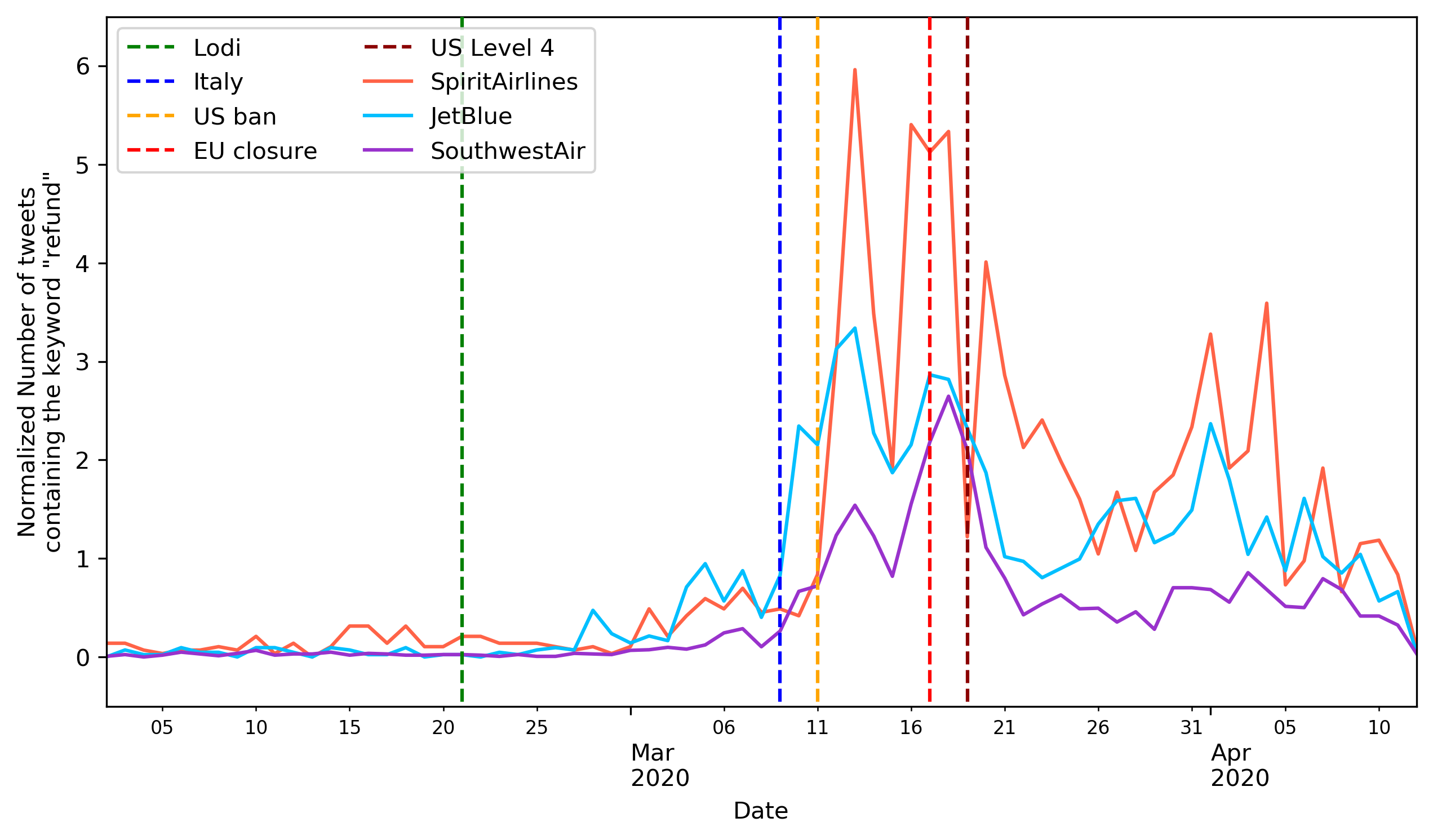}
\caption{From passengers of low-cost airlines}
\label{fig:kw_refund_pax_lowcost}
\end{subfigure}
\caption{Number of occurrences of the keyword "refund" in tweets written by passengers normalized by the number of total passengers per carrier over the year 2018 using BTS data \cite{BTS2018carrier}}
\label{fig:kw_refund_pax}
\end{center}
\end{figure}

Figure\,\ref{fig:kw_refund_pax} shows the evolution of the number of occurrences of the keyword "refund" in tweets written by passengers from March 1\st\, 2020 to April 12\nth\,2020 for the same seven US airlines.

Figure\,\ref{fig:kw_refund_pax} shows the evolution of the normalized number of occurrences of the keyword "refund" in tweets written by passengers from February 1\st\,2020 to April 12\nth\, 2020 for the same seven US airlines using the same normalization process.

From a passenger perspective, the situation linked to the "refund" keyword is similar to the situation linked to the "cancel" keyword but at a lower proportion. Figure\,\ref{fig:kw_refund_pax_major} shows that legacy airlines all have a steep increase in the number of occurrences of the keyword "refund" in passenger tweets at the announcement of Italy's lockdown and then a very slow decrease, with Alaska Airlines have an anticipated spike at the beginning of March 2020.
Figure\,\ref{fig:kw_refund_pax_lowcost} shows that Southwest Airlines increase in the occurrences of the keyword "refund" is still lower and resorbs faster than the other low-cost carriers and that the spike for Spirit Airlines passengers starts only at the announcement of the US travel ban.

\begin{figure}[h!t]
\begin{center}
\begin{subfigure}[b]{.49\textwidth}
\includegraphics[width=\textwidth]{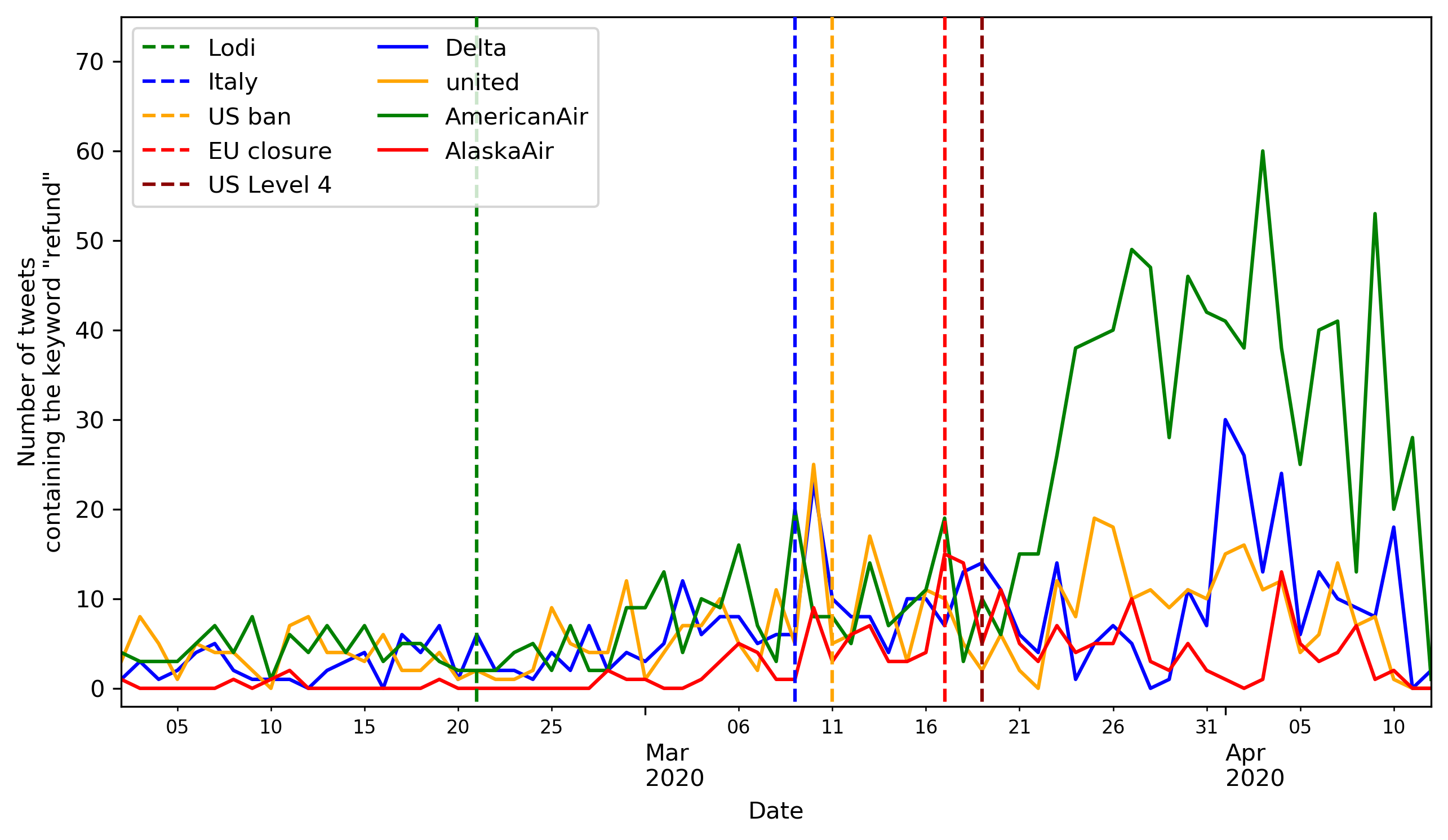}
\caption{From customer service of legacy airlines}
\label{fig:kw_refund_cie_major}
\end{subfigure}\hfill
\begin{subfigure}[b]{.49\textwidth}
\includegraphics[width=\textwidth]{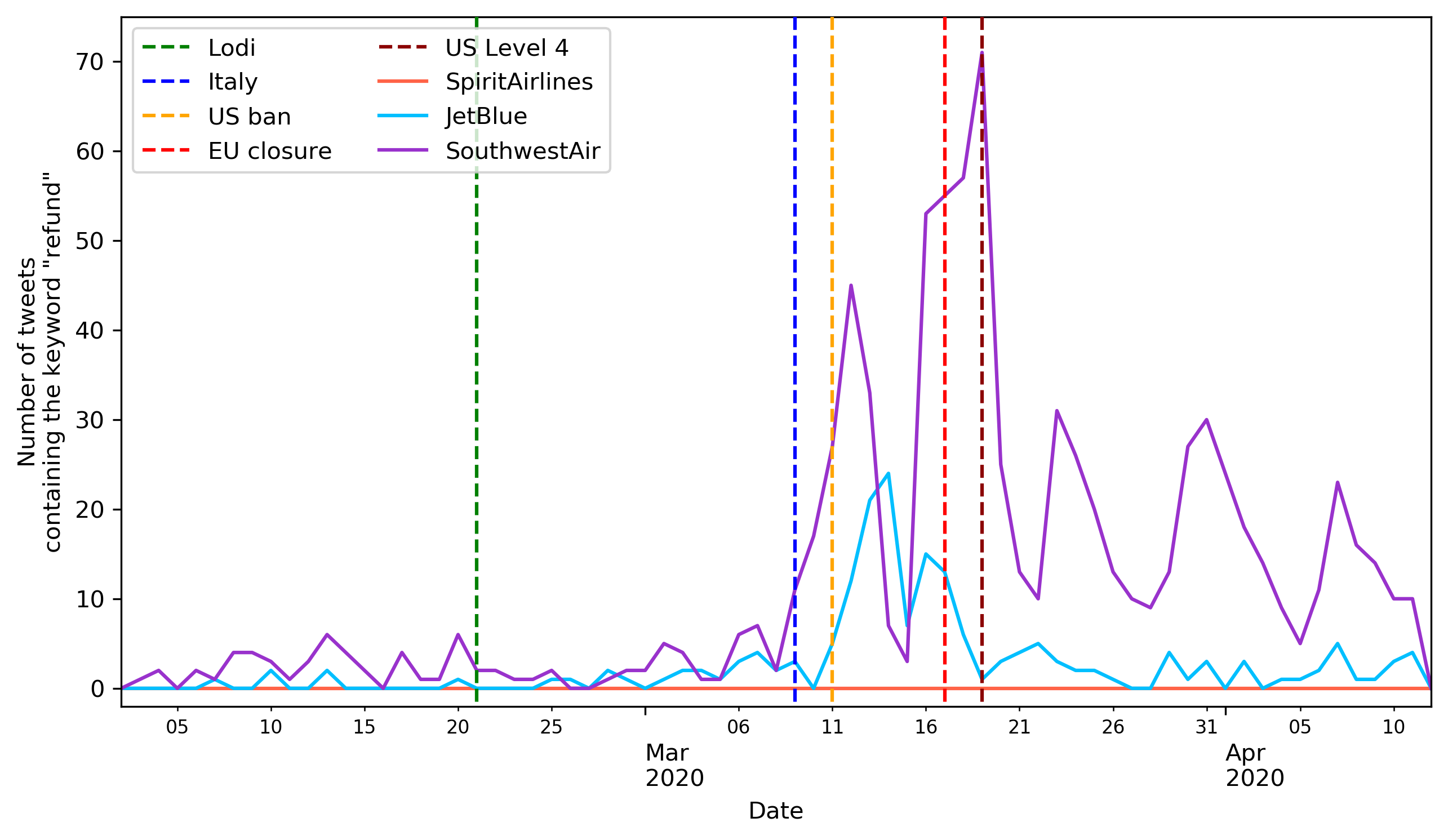}
\caption{From customer service of low-cost airlines}
\label{fig:kw_refund_cie_lowcost}
\end{subfigure}
\caption{Number of occurrences of the keyword "refund" in tweets written by airline customer services}
\label{fig:kw_refund_cie}
\end{center}
\end{figure}

Figure\,\ref{fig:kw_refund_cie} shows the evolution of the number of occurrences of the keyword "refund" in tweets written by airline customer services from February 1\st\, 2020 to April 12\nth\,2020 for the same seven US airlines.

Figure\,\ref{fig:kw_refund_cie_major} shows the evolution of the number of occurrences of the keyword "cancel" within tweets of customer service of the four considered legacy airlines. The initial increase is similar than for the keyword "cancel" (Figure\,\ref{fig:kw_cancel_cie_major}), however there is then a second increase towards the end of March 2020, this increase being most visible within the tweets written by American Airlines.
From a low-cost carrier perspective, Figure\,\ref{fig:kw_refund_cie_lowcost} illustrates the same idiosyncracies as in Figure\,\ref{fig:kw_cancel_cie_lowcost}: Two spikes around the US announcements for Southwest Airlines, this time with higher fluctuations afterwards, one major spike on March 13\nth\,2020 for JetBlue and not even one occurrence of the keyword "refund" over the whole period for Spirit Airlines.

The same two metrics associated to the "cancel"-related Twitter situation presented in Section\,\ref{sec:kw_cancel}, i.e. the quality response score and the quantity response score, can be used for this "refund"-related Twitter situation.
Table\,\ref{tab:kw_refund_ranking} presents these two proposed metrics in the case of the keyword "refund" using the same predefined threshold of 1 for delimiting a Twitter situation.

\begin{table}[ht]
\begin{center}
\caption{Airline ranking based on the proposed keyword metrics applied to the period of March 1\st\,2020 to April 11\nth\,2020 with the keyword "refund"}
\label{tab:kw_refund_ranking}
\begin{tabular}{|c||c|c||c|c|}
\hline
\bfseries Rank & \bfseries Airline & \bfseries Quality & \bfseries Airline & \bfseries Quantity \\ \hline
1&SouthwestAir&8&SouthwestAir&43.25\\
2&AlaskaAir&25&AmericanAir&26.65\\
3&JetBlue&25&Delta&10.89\\
4&SpiritAirlines&27&united&9.48\\
5&Delta&28&AlaskaAir&5.92\\
6&AmericanAir&31&JetBlue&5.28\\
7&united&33&SpiritAirlines&0.00\\

\hline
\end{tabular}
\end{center}
\end{table}

\section{Discussion \& Conclusion}
\label{sec:concl}
\subsection{Summary of the proposed metrics}
In order to complement existing flight-centric metrics, which are not broadly and readily available, this paper proposed to consider seven different passenger-centric metrics based on passenger-generated data, three of them focusing on airports and the other four metrics focusing on airlines. They are regrouped and listed here.

\begin{paxmetric2}
\label{pax2:immigration}
The \textbf{immigration quality score} for an airport of entry is defined as the correlation between the daily average wait time for passengers at its immigration service and the daily average immigration load factor of the airport over a given period.
\end{paxmetric2}

\begin{paxmetric2}
\label{pax2:prop60}
The weekly \textbf{airport visitor efficiency score} for an airport is defined as the weekly proportion of airport visitors spending less than 60 minutes at an airport.
\end{paxmetric2}

\begin{paxmetric2}
\label{pax2:prop240}
The weekly \textbf{airport visitor slugginess score} for an airport is defined as the weekly proportion of airport visitors spending more than 240 minutes at an airport.
\end{paxmetric2}

\begin{paxmetric2}
\label{pax2:empathy}
The airline \textbf{empathy} score is defined as the correlation between the evolution of the average mood expressed by passengers in their tweets and the evolution of the average mood expressed by the airline customer service in their tweets.
\end{paxmetric2}

\begin{paxmetric2}
\label{pax2:sent_gap}
The airline \textbf{sentiment gap} measures the average difference between the average mood expressed by passengers and the average mood expressed by airlines.
\end{paxmetric2}

The following two proposed metrics can be used with selected keywords, e.g. "cancel" and "refund".
\begin{paxmetric2}
\label{pax2:kw_quality}
The keyword-related Twitter situation \textbf{quality response} score of an airline is the time needed for the airline to bring the normalized number of tweets containing the specific keyword back to a predefined level.
\end{paxmetric2}

\begin{paxmetric2}
\label{pax2:kw_quantity}
The keyword-related Twitter situation \textbf{quantity response} score of an airline is calculated by integrating the number of tweets written by the airline and containing the keyword over the number of days associated to the situation.
\end{paxmetric2}

\subsection{Discussion}
These proposed passenger-centric metrics were built using data available, either in real-time or due to the exceptional circumstances, and can still be further tuned to meet the expectations of both federal agencies and passengers. Several limitations and possible improvements should be noted here for a better understanding of these proposed metrics.

The data used to estimate the number of visitors at airports and the proportion of time spent per airport location was graciously provided by SafeGraph in order to better understand the COVID-19 situation, and is not usually as easily available. In order to implement the associated metrics, agreements should be held between the different data providers and the group in charge of such metrics. Furthermore, an analysis of the categories of person most likely to be within the gathered data should be undertaken to better tune the final score.

The proposed Twitter related metrics have the unique advantage, among the presented datasets, that they can easily be updated on an hourly basis if needed. They also enable each passenger and airline to actively be able to influence their scores. They measure however essentially the communication quality and quantity between airlines and passengers, and should therefore still be complemented with traditional flight-centric measures for completeness.

This study focused on the effects of the travel restriction measures linked to a major disruption taking its course over an important number of days and tailored the proposed metrics for this timespan. Future studies could also investigate into the adaptation of some of these proposed passenger-centric metrics to measure effects on a smaller scale, e.g. over a single day or a few hours.

%
% end content
%%%%%%

%%%
% If any
%
\section*{Acknowledgments}
The authors would like to thank Nikunj Oza from NASA-Ames, the French \emph{École Nationale de l'Aviation Civile} and the King Abdullah University of Science and Technology for their financial support, as well as SafeGraph for making their data available for this study.

The authors would also like to deeply thank all workers and researchers associated in the fight against the COVID-19 pandemic, with a special thought to health-care workers and providers.
%
%%%

%%%
% Biblio
%
\bibliography{covid.bib}
%
%%%

\end{document}